\documentclass[12pt,preprint]{aastex}
\begin{document}

\parindent=1.0cm

\title{The Stellar Disk of M81\altaffilmark{1}}

\author{T. J. Davidge}

\affil{Herzberg Institute of Astrophysics,
\\National Research Council of Canada, 5071 West Saanich Road,
\\Victoria, BC Canada V9E 2E7\\ {\it email: tim.davidge@nrc.ca}}

\altaffiltext{1}{Based on observations obtained with the Megaprime/MegaCam,
a joint project of the CFHT and CEA/DAPNIA, at the Canada-France-Hawaii Telescope,
which is operated by the National Research Council of Canada, the Institut National
des Sciences de l'Univers of the Centre National de la Recherche Scientifique of France,
and the University of Hawaii.}

\begin{abstract}

	Wide-field images obtained with the 3.6 meter Canada-France-Hawaii Telescope 
are used to investigate the spatial distribution and photometric 
properties of the brightest stars in the disk of M81 (NGC 3031). With the exception 
of the central $\sim 2$ kpc of the galaxy and gaps between CCDs, the survey is 
spatially complete for stars with $i' < 24$ and major axis distances of 18 kpc. 
A more modest near-infrared survey detects stars with $K < 20$ over 
roughly one third of the disk. Bright main sequence (MS) stars 
and RSGs are traced out to galactocentric distances of at least 18 kpc. 
The color of the red supergiant (RSG) locus suggests 
that Z = 0.008 when R$_{GC} > 6$ kpc, and such a radially uniform RSG metallicity is 
consistent with [O/H] measurements from HII regions. The density of bright MS stars and RSGs 
drops when R$_{GC} < 4$ kpc, suggesting that star formation in the inner disk was 
curtailed within the past $\sim 100$ Myr. The spatial distribution of bright MS stars tracks 
emission at far-ultraviolet, mid- and far-infrared wavelengths, although tidal 
features contain bright MS stars but have little or no infrared flux. The 
specific frequency of bright MS stars and RSGs, normalized to $K-$band integrated brightness, 
increases with radius, indicating that during the past $\sim 30$ Myr 
the specific star formation rate (SSFR) has increased with increasing radius. Still, the SSFR 
of the M81 disk at intermediate radii is consistent with that expected for an isolated galaxy 
as massive as M81, indicating that the star formation rate in the disk of M81 has not been 
markedly elevated during the past few tens of millions of years. The stellar content of the 
M81 disk undergoes a distinct change near R$_{GC} \sim 14$ kpc; the $K-$band light profile, 
which is dominated by old and intermediate age stars, breaks downward at this radius, 
whereas the density profile of young stars flattens, but does not break downwards. Thus, 
the luminosity-weighted mean age decreases with increasing radius in the outer regions of the 
M81 disk.

\end{abstract}


\section{INTRODUCTION}

	The formation of spiral galaxy disks is likely a long-term process 
that continues to the present day. Bekki \& Chiba (2001) model the formation of a 
Galaxy-like system, and find that while a central spheroid forms from the 
merger of dominant clumps $\sim 8$ Gyr in the past, a stable gas disk only forms $\sim 
2$ Gyr later. Disks that formed at moderate to high redshift may continue to grow to the 
present-day, as fresh material is accreted (e.g. Trujillo \& Pohlen 
2005; Barden et al. 2005). Naab \& Ostriker (2006) model the 
infall of material onto the Galactic disk, and find that the disk 
grows outward with time, with a rate of growth at the present day of $\sim 1$ kpc 
Gyr$^{-1}$. That disk size increases with time is consistent with observations of 
galaxies spanning a range of redshifts (Trujillo \& Aguerri 2004; Trujillo \& Pohlen 2005). 

	Large-scale radial trends are likely imprinted in disks during their assembly. 
While secular processes will blur these trends on timescales of many crossing times 
(e.g. Sellwood \& Binney 2002; Roskar et al. 2008), mergers and galaxy-galaxy interactions 
may have an even greater and more rapid impact on the distribution of stars and gas. 
There is a high merger rate among spiral galaxies at intermediate redshifts (Hammer 
et al. 2005) and, depending on the orbital geometry, the accretion of a minor satellite 
may have a profound impact on the structural properties 
of disks, causing rings, warps, and a vertical morphology that can be loosely
interpreted in the context of thin and thick disk components (e.g. Read et al. 2008; 
Kazantzidis et al. 2008). Galaxy-galaxy interactions may drive gas into the 
central regions of galaxies (e.g. Mihos \& Hernquist 1994), resulting in 
centrally-concentrated star-forming activity, and the choking of 
star formation at large radii, as may have happened in M82 (Davidge 2008a). 

	While major mergers will have a large impact on the properties of the host 
systems, these may not permanently obliterate disks if the merging systems are 
gas-dominated, as gas disks can re-form after a merger (e.g. Robertson et al. 
2006; Governato et al. 2007). Based largely on the angular momentum 
of disks and the chemical composition of stars at large radii, 
Hammer et al. (2007) argue that the majority of nearby spirals 
experienced major mergers within the past $\sim 6$ Gyr, and 
that the Galaxy has apparantly escaped such activity. While the main 
period of mass assembly in the Galaxy may then have terminated some 8 - 10 Gyr in the past 
(e.g. Zentner \& Bullock 2003; Bullock \& Johnston 2005), the star-forming 
history of the solar neighborhood shows evidence of a possible accretion event during 
intermediate epochs (Cignoni et al. 2006), suggesting that the Galactic disk 
may not have completely escaped interactions with companions.

	Located at a distance of only 3.9 Mpc, the SA(s)ab galaxy M81 is an 
important laboratory for probing the impact of interactions on the disk of 
a large spiral galaxy. Dynamical arguments (Yun, Ho, \& Lo 1994) and the stellar content 
of M82 (e.g. de Grijs, O'Connell, \& Gallagher 2001; Mayya et al. 2006) 
indicate that M81 and M82 interacted a few hundred Myr ago, 
when M82 appears to have passed through the disk of M81 (Yun et al. 1994). Streams of HI
link M81, M82, and NGC 3077 (e.g. Brouillet et al. 1991; Yun et al. 1994; Boyce et al. 
2001). The debris field between these galaxies is an area of recent star formation (e.g. 
de Mello et al. 2008), and diffuse ensembles of young stars are 
present (Durrell et al 2004; Davidge 2008b).

	Previous studies of the stellar content of M81 have explored only a 
small part of the disk. Hughes et al. (1994) discuss two 
WFPC2 fields that sample the disk of M81 at intermediate radii. Their CMD has a broad 
blue plume that contains a mix of main sequence (MS) stars and blue supergiants (BSGs) 
and a red supergiant (RSG) sequence; the blue and red plumes both peak near M$_V \sim -7.5$. 
Tikhonov, Galazutdinova, \& Drozdovsky (2005) probe the stellar content 
of M81 in 6 WFPC2 fields, and resolve red giant branch 
(RGB) stars in some of these. Using the RGB-tip brightness, they estimate 
the distance of M81 to be 3.85 Mpc, and this distance is adopted for M81 throughout 
this study. Tikhonov et al. (2005) also detect radial gradients in the properties 
of the RGB population that they attribute to a metallicity gradient. Based on four fields 
that sample the outer disk, they find that RGB stars have a mean metallicity [Fe/H] 
$= -0.65 \pm 0.04$.

	There is evidence that the interaction with M82 had an impact on the recent star-forming 
history of M81. Chandar, Tsvetanov, \& Ford (2001) find a population of compact blue star 
clusters in M81, the formation of which appeared to have commenced $\sim 600$ Myr in the past. 
Clusters of this nature are typically associated with elevated levels of star-forming 
activity, and the youngest of these have ages $\sim 6$ Myr.
Swartz et al. (2003) find a population of x-ray sources in the central 2 kpc 
of M81 that probably belongs to the bulge. With an estimated age $\sim 400$ Myr, these 
sources appear to be the remnant of an elevated episode of centrally-concentrated 
star formation. Davidge (2006a) examines the distribution of 
RSGs in four disk fields at intermediate radii, and finds evidence 
for spatial variations in the star-forming history during the past $\sim 25$ Myr. 
These data suggest that star formation was distributed over a larger fraction of 
the disk 25 Myr in the past than at the present day, and that 
the northern portion of the disk has been a site of long-term star-forming activity. 
Most recently, Williams et al. (2008) discuss deep ACS images of a field in the outer 
disk of M81. When averaged over Gyr time scales, the star formation rate (SFR) in this 
field stayed roughly constant for a large fraction of the age of the Universe. 
However, a few hundred Myr ago the SFR declined, in marked contrast with 
the inner regions of the galaxy. A spiral arm passes 
through the Williams et al. (2008) field, and the SFR in the arm increased 
during the past few tens of Myr. Williams et al. (2008) conclude that the stars in the 
outer disk of M81 experienced rapid early enrichment, and that the 
stars that formed in the past 0.1 Gyr typically have [M/H] $\sim -0.4$. 

	The studies discussed above indicate that there have been spatial and temporal 
variations in the star-forming history of M81 during the past few hundred Myr. 
Arguably of greatest interest in the context of galaxy-galaxy interactions is 
the increase in the central SFR during this time, and the corresponding drop in the  
SFR of the outer disk, likely due to the movement of gas in the galaxy. 
Measuring the ages and metallicities of stars across the disk of M81 will help establish the 
chronology of these events, and the timescales over which gas has moved 
since the interaction. A census of the brightest 
blue and red stars in M81 is a modest but logical starting point to investigate 
spatial variations in the star-forming history 
during the past $\sim 100$ Myr. Stellar structure models also suggest 
that the color of the RSG sequence is sensitive to metallicity, and so it can be used 
to gauge the metallicities of the youngest stars, and investigate possible radial trends.

	Efforts to study the stellar content of the outer 
disk of M81 are complicated by contamination from tidal debris. Some tidal companions 
(e.g. Makarova et al. 2002; Sabbi et al. 2008) are relatively dense stellar aggregates, and 
stars from these objects can be culled from the data with relative ease. However, diffuse 
collections of young stars that are probably in the process of tidal disruption have 
also been identified (e.g. Durrell et al. 2004; Davidge 2008b). While the objects 
identified by Durrell et al. (2004) and Davidge (2008b) are located 
outside of the area considered in this paper, the identification and 
removal of stars that might belong to such low density structures of this type 
that are closer to M81 is problematic. Still, a survey that covers the 
entire disk of M81 can be used to suppress the impact of 
such structures by azimuthally averaging stellar properties over the entire disk.

	In the present study, the wide-field MegaCam and WIRCam imagers 
on the 3.6 meter Canada-France-Hawaii Telescope (CFHT) are used to investigate the 
photometric properties and spatial distributions of the 
brightest stars in the M81 disk. With the exception of the crowded central regions 
of the galaxy and gaps between detector elements, the MegaCam data cover 
the entire disk of the galaxy and the debris field between M81 and M82, sampling the 
brightest MS stars, BSGs and RSGs. The WIRCam data covers 
much of the south west quadrant of the galaxy, detecting the brightest 
RSGs and asymptotic giant branch (AGB) stars in this part of the galaxy.

	The paper is structured as follows. The observations, along with the procedures used to 
reduce the raw data and make the photometric measurements, are described in \S 2. 
The photometric properties of the brightest blue and red stars 
are discussed in \S 3, while the spatial distribution of the brightest stars is examined 
in \S 4. A summary and discussion of the results follows in \S 5. In Appendix A the 
old age estimate for M81 that is determined from integrated light studies is reconciled with 
the evidence for an elevated level of star formation in the central regions of the galaxy, 
while in Appendix B near-infrared measurements of spectroscopically confirmed globular 
clusters are presented.

\section{OBSERVATIONS \& REDUCTIONS}

\subsection{MegaCam}

	Images of a one degree$^2$ field that is centered midway between M81 and M82 were
recorded with MegaCam (Boulade et al. 2003) through $r', i'$ and $z'$ filters on the
night of October 23 UT 2006. The detector in Megacam is a mosaic of
thirty six $2048 \times 4612$ pixel$^2$ CCDs, with 0.185 arcsec pixel$^{-1}$ sampling.
A four-point square dither pattern was used to assist with the identification of bad 
pixels and cosmic rays. Four 300 sec exposures were recorded in $r'$ and $i'$,
while four 500 sec exposures were recorded in $z'$. Stars have 0.7 -- 0.8 arcsec
FWHM in the final processed images, depending on the filter.

	Instrumental and environmental signatures were removed from the MegaCam data 
with the CFHT ELIXIR pipeline, which performs bias subtraction,
flat-fielding, and the subtraction of a fringe frame. 
The ELIXIR-processed images were aligned, stacked, and then trimmed 
to the area that is common to all exposures. These data were used previously to examine 
the stellar content in and around M82 (Davidge 2008a,c), and search for diffuse 
stellar groupings in the debris field between M81 and M82 (Davidge 2008b).

\subsection{WIRCam}

	The western disk of M81 was imaged through $J, H$ 
and $Ks$ filters with WIRCam (Puget et al. 2004) on the night of February 4 UT 2007. 
The field is centered on Ho IX. The detector in 
WIRCam is a mosaic of four $2048 \times 2048$ HgCdTe arrays, that together 
image a $21 \times 21$ arcmin$^2$ field with 0.3 arcsec pixel$^{-1}$ sampling. 

	A square-shaped dither pattern was used during data acquisition. A set of $20 \times 
45$ sec exposures was recorded in $J$, while $120 \times 15$ sec exposures were recorded 
in $H$, and $160 \times 15$ sec exposures were recorded in $Ks$. The $J$ images were obtained
over one dither cycle (i.e. five 45 sec exposures were recorded per dither position),
while the $H$ and $Ks$ images were obtained over two complete dither cycles.
Stars in the final images have 0.8 arcsec FWHM.

	The initial processing of these data was done with the CFHT I`IWI pipeline, 
and this consisted of dark subtraction and flat-fielding. A calibration frame
to remove interference fringes and thermal emission artifacts
was constructed by median-combining all I'IWI-processed images 
obtained for this program, including those of M82 
(Davidge 2008a). A low-pass clipping algorithm was applied 
to suppress stars and galaxies in the combined images, and the resulting calibration frames 
were subtracted from the flat-fielded data. The de-fringed images were aligned,
stacked, and then trimmed to the area common to all exposures.

\subsection{Photometric Measurements}

	The photometric measurements were made with the point spread function (PSF) 
fitting routine in ALLSTAR (Stetson \& Harris 1988). The PSFs were constructed 
iteratively using the DAOPHOT (Stetson 1987) PSF task, with 
contaminating objects close to PSF stars being subtracted using progressively improved PSFs. 
Each PSF was typically constructed from 80 -- 100 stars. 
Image quality varies by small amounts across the $1 \times 1$ degree MegaCam field. 
To mitigate the impact of these variations, the MegaCam images were divided into six 30 
arcmin $\times$ 20 arcmin panels, and separate PSFs were generated for each 
panel. An inspection of star-subtracted images showed that the 
PSFs provided acceptable fits over each panel. The image quality is more stable across the 
smaller WIRCam field, and a single PSF was constructed in each filter 
using stars from all four detector elements.

	The photometric catalogues were culled to reject objects
with large uncertainties in their measurements. Objects for which the fit error 
computed by ALLSTAR, $\epsilon$, exceeded $\pm 0.3$ magnitudes were rejected to 
remove sources near the faint limit of the data, where photometry is problematic. 
In addition, $\epsilon$ tends to rise monotonically towards fainter magnitudes 
for the majority of objects, and there are some objects for which $\epsilon$ is 
markedly higher than the norm at a given brightness. Such objects were 
also removed, and these tend to be either (1) galaxies, (2) multi-pixel 
cosmetic defects, and/or (3) in the crowded central regions of M81.

	Photometric zeropoints that are measured from standard stars that 
are observed as part of each MegaCam run are placed in MegaCam data headers during 
ELIXIR processing. The photometric calibration of the M81 Megacam data used the zeropoints 
measured in October 2006. As for WIRCam, the calibration used 
zeropoints that are posted on the CFHT website, which were computed from 
standard star observations made in February 2007.

	The photometric calibration was checked using published 
photometric catalogues. The MegaCam photometric measurements were compared with 
entries in the Sloan Dgital Sky Survey (SDSS) Data Release 7 (Adelman-McCarthy et al. 2009, 
in preparation). The differences between the MegaCam and SDSS measurements for sources 
with $i'$ between 17 and 19 are $\Delta i' = 0.025 \pm 0.012$, $\Delta (r'-i') = 
-0.001 \pm 0.015$, and $\Delta (i'-z') = 0.013 \pm 0.020$, where the uncertainties are 
the standard errors of the mean. The WIRCam calibration 
was checked using entries with $K < 14.5$ in the 2MASS Point Source 
Catalogue (Cutri et al. 2003). The average differences between measurements in the WIRCam 
and 2MASS systems are $\Delta K = -0.03 \pm 0.05$, $\Delta (J-K) = 0.01 \pm 0.06$, 
and $\Delta (H-K) = -0.01 \pm 0.06$.

	Sample completeness and the photometric scatter due to photon noise and crowding 
were assessed from artificial star experiments. Artificial stars were 
assigned colors that follow the main locus of stars in the M81 CMDs, and their brightnesses 
were measured using the same procedures that were applied to the actual data. 
Artificial stars were considered to be recovered only if they were detected in at least two 
bandpasses, after applying the $\epsilon$-based rejection critera described earlier. 

	The artificial star experiments indicate that 50\% completeness 
is encountered near $i' = 24.5$ and $K = 20$ throughout most of the M81 disk. 
Stellar density increases with decreasing distance from the center of M81, and so 
incompleteness sets in at brighter magnitudes in the central 4 kpc of 
M81. The WIRCam data is less susceptible to changes in stellar density because there is 
greater contrast in the infrared between the brightest red stars 
and the bluer unresolved stellar body than at visible wavelengths. 
The artificial star experiments further indicate that blends may account 
for a significant fraction of detected objects in the MegaCam and WIRCam 
data at magnitudes where the completeness fraction $< 50\%$. 

\section{RESULTS: THE CMDs}

\subsection{The Morphology of the CMDs}

\subsubsection{The MegaCam Data}

	The $(i', r'-i')$ and $(z', i'-z')$ CMDs of sources in various annuli 
are shown in Figures 1 and 2. The distance interval specified in each panel is 
in the M81 disk plane and assumes that the disk is inclined at 59.3 degrees, 
based on the ellipticity of M81 in 2MASS images (Jarrett et al. 2003). 
There are stellar ensembles in the peripheral regions of M81 that are likely tidal in origin 
(e.g. Sun et al. 2005, Durrell et al. 2004, Davidge 2008b). The most obvious tidal 
contamination comes from Ho IX, and sources that belong to Ho IX have 
been culled from the data plotted in Figures 1 and 2. 

	The CMDs of sources with R$_{GC}$ between 6 and 14 kpc have similar morphologies. 
Given that stellar density changes with R$_{GC}$, then this 
radial stability of the CMDs suggests that crowding does not affect the 
photometry of the brightest stars in this radial interval. The dominant 
feature in the $(i', r'-i')$ CMDs of sources with R$_{GC} < 14$ kpc is a concentration of 
objects that have $i' > 22.5$ and $r'-i'$ between 0 and 1.5. These objects are a 
combination of RSGs with ages in excess of $\sim 50 - 100$ Myr and the brightest AGB stars. 
The mixing of stars with such different evolutionary states and progenitor properties 
produces an amorphous cloud-like structure in the lower portions of the CMDs.

	Bright RSGs form a sequence that rises out of the concentration of AGB 
stars and fainter RSGs in the CMDs of objects with R$_{GC}$ between 4 and 12 kpc. 
The RSG sequence contains stars with $r'-i'$ between 
0.5 and 1.0, and $i'-z'$ between 0 and 0.2. The diminished 
impact of line blanketing on the photometric properties of RSGs in 
the $(z', i'-z')$ CMDs is clearly evident when compared with the $(i', r'-i')$ CMDs, as 
RSGs form a near-vertical sequence in the $(z', i'-z')$ CMDs. This 
tighter morphology aids efforts to trace RSGs out to large radii, and 
a RSG sequence is seen in the 12 -- 14 kpc $(z', i'-z')$ CMD, but 
not in the $(i', r'-i')$ CMD of the same radial interval. 

	A prominant population of objects with $r'-i' < 0$ is seen in the $(i', r'-i')$ CMDs 
of most radial intervals. Although less pronounced because of the redder wavelength 
coverage, a corresponding sequence of blue objects is also present in the $(z', i'-z')$ CMDs. 
Given that the number density of background galaxies and foreground stars with $r'-i' 
< 0$ and $i'$ between 20 and 24 is modest (e.g. Davidge 2008b), then the blue objects 
in the CMDs are almost certainly MS stars, BSGs, and unresolved compact young star clusters 
that belong to M81 or the surrounding tidal debris field.

	The number of bright MS stars is comparatively modest in the 2 -- 4 kpc interval, 
suggesting that the SFR in this part of M81 during recent epochs was 
lower than at larger radii. Gordon et al. (2004) use far-infrared 
flux measurements obtained from {\it Spitzer} MIPS data to 
investigate the SFR throughout the disk of M81, and the 2 -- 4 kpc interval overlaps 
with the region that they refer to as the `inner ring'. Gordon et al. (2004) find that 
the specific SFR (SSFR) in the inner ring is near the lower limit of what is seen 
throughout the entire disk, in qualitative agreement with the modest number of bright 
MS stars in the CMDs of this part of M81.

	Gordon et al. (2004) found that the SFRs estimated for M81 from UV and H$\alpha$ 
emission are systematically lower than those from $24\mu$m emission 
due, at least in part, to star-forming areas that are 
obscured by dust at visible wavelengths. Much of the 
$24\mu$m emission in M81 is concentrated in the 4 -- 6 kpc annulus, 
and the MS and RSG plumes in the CMDs of this annulus 
are broader than at larger radii. This suggests that there may be a larger amount of 
dust mixed with young stars and along the line of sight 
in the 4 -- 6 kpc annulus than at larger radii. 

	The majority of objects in the 16 -- 18 kpc CMDs are foreground stars and background 
galaxies. Foreground Galactic stars form a diffuse population of objects with $r'-i' > 0$ 
in the $(i', r'-i')$ CMDs, while background galaxies have relatively red 
colors and dominate over foreground stars when $i' > 22$. 
Foreground stars and background galaxies make progressively larger contributions to the CMDs 
as R$_{GC}$ increases because as one moves to larger R$_{GC}$ then 
(1) the density of sources that belong to M81 diminishes, and (2) larger areas on the 
sky are sampled in each annulus. 

\subsubsection{The WIRCam Data}

	The $(K, J-K)$ and $(K, H-K)$ CMDs constructed from the 
WIRCam observations, which sample roughly one third of the M81 disk, are shown in 
Figures 3 and 4. As in the preceeding section, stars that belong to Ho IX have been 
excised from the CMDs at large R$_{GC}$. The most prominent feature in the near-infrared 
CMDs is the cloud of AGB stars and older RSGs that have $K > 19$, or M$_K > -9$. 
This feature can be traced out to R$_{GC} \sim 16$ kpc in the near-infrared CMDs. 

	The most luminous RSGs and AGB stars form a vertical sequence centered 
near H--K $\sim 0.2$ and J--K $\sim 1$, that has $K > 17$, or 
M$_K > -11$. The bright RSG/AGB sequence is 
seen out to R$_{GC} = 16$ kpc. Only a modestly populated bright RSG sequence 
is seen in the $2 - 4$ kpc CMDs, in agreement 
with the low number of RSGs in the MegaCam CMDs of this radial interval.

	The majority of objects in the $16 - 18$ kpc WIRCam CMDs are 
foreground stars and background galaxies. The foreground stars occupy the sequence with 
$J-K \sim 0.7$ and $H-K \sim 0.2$ that has $K < 18$. Background galaxies 
populate the diffuse cloud of objects with $K > 17$ that is centered 
near $H-K \sim 0.8$ and $J-K \sim 1.6$, which are the approximate 
colors of normal galaxies at intermediate redshifts (e.g. Davidge 2007).

\subsection{Comparisons with Isochrones}

\subsubsection{The M81 Disk at Intermediate and Large Radii}

	The CMDs of stars with R$_{GC}$ between 8 and 10 kpc are 
representative of the disk of M81 at intermediate radii, and in Figure 5 the $(M_{i'}, r'-i')$ 
and $(M_{z'}, i'-z')$ CMDs of this interval are compared with isochrones 
from Girardi et al. (2004). Isochrones with log(t$_{yr}) = 7.0$, 7.5, 
8.0, and 8.5 are shown for metallicities Z = 0.008 and Z = 0.019. The isochrones 
were constructed from a compilation of stellar evolution 
models generated by the Padova group, and those shown in Figure 5 rely 
on the evolutionary sequences discussed by Salasnich et al. (2000), with 
a thermally-pulsing AGB component as described by Marigo \& Girardi (2001). The models 
include convective overshooting and metallicity-dependant mass loss in the input physics.

	The blue envelope of points on the $(i', r'-i')$ and $(z', i'-z')$ 
CMDs is well matched by the log(t$_{yr}$) = 7.0 isochrones, indicating that there has 
been very recent star formation in the 8 -- 10 kpc interval. A modest number 
of stars fall above the log(t$_{yr}$) = 7.0 MS turn-off on both CMDs. While the majority of 
these are probably foreground stars, the possibility that some of these are 
either stars younger than log(t$_{yr}$) = 7 or blends can not be discounted with these data. 

	The brightest RSGs define a distinct sequence 
in visible and near-infrared wavelength CMDs, and the isochrones predict that 
the color and shape of the RSG sequence is sensitive to metallicity. The 
Z = 0.008 models match the $r'-i'$ and $i'-z'$ colors of the RSG 
sequence in the 8 -- 10 kpc interval, whereas the Z = 0.019 models 
predict $r'-i'$ and $i'-z'$ colors that are a few tenths of a mag redder
than observed. In addition, the Z = 0.019 RSG sequences are more curved 
at the upper end than the Z = 0.008 sequences, and in this regard the Z = 0.008 isochrones 
again better match the observations than the Z = 0.019 
isochrones. Thus, both the color and shape of the RSG sequences 
in the $(i', r'-i')$ and $(z', i'-z')$ CMDs are consistent with Z = 0.008. 
This is consistent with the metallicity calculated by Williams et al. (2008) 
for young stars in the outer disk of M81. RSGs in the disk of M81 and 
the disks of NGC 247 and NGC 2403 have similar metallicities (Davidge 2006b; 2007). 

	The reader is cautioned that the calibration and metallicity-sensitivity of the 
portions of isochrones that track the post-MS stages of massive star evolution 
are uncertain, and the impact of mass loss, convective overshooting, and rotation 
on models of the RSG phase of evolution can be substantial. 
It is encouraging that the isochrones used here have been compared with 
the CMDs of fiducial star clusters with good results (e.g. Girardi et al. 2004; Bonatto, 
Bica, \& Girardi 2004). However, while well-studied, the fiducial clusters are 
either old (e.g. Girardi et al. 2004), or have only modest numbers of highly evolved stars 
(Bonatto et al. 2004), so that the comparisons effectively probe 
only stars that have experienced modest amounts of evolution.

	The ratio of blue to red supergiants is a fundamental diagnostic of 
massive star models, and the inability of early models to reproduce this ratio (Langer \& 
Maeder 1995), was attributed to not including rotation in the input physics
(e.g. Maeder \& Meynet 2001; Dohm-Palmer \& Skillman 2002). This is a short-coming of 
the isochrones used here. Moreover, it is well-established that the mean metallicity 
of galaxies varies with system mass. It is thus worth noting that the calibration in 
Figure 5b of Asari et al. (2007) indicates that the metallicity of stars in M81 (M$_K \sim 
-24$) should be $\sim 0.2$ dex higher than of those in NGC 2403 (M$_K \sim -22$), whereas 
the mean colors of the RSG in these galaxies are similar. For comparison, the [O/H] values 
of M81 and NGC 2403 are offset by a few tenths of a dex (Figure 8 of Zaritsky et al. 1994).

	The WIRCam observations of the 8 -- 10 kpc interval are compared with 
Z = 0.008 and Z = 0.019 isochrones from Girardi et al. (2002) in Figure 6. 
The isochrones have ages log(t$_{yr}$) = 7.5, 8.0, 8.5, and 9.0, and are based on the 
Salasnich et al. (2000) evolutionary sequences. Because of the diminished impact of 
line blanketing on photometric properties in the near-infrared, coupled with the 
reduced sensitivity to changes in effective temperature for all but the coolest stars, the AGB 
sequences of the oldest isochrones in the near-infrared are closer to vertical than at 
visible wavelengths. The isochrones suggest that stars evolving near the AGB-tip 
that are older than log(t$_{yr}$) = 9.0 are resolved in the WIRCam images.
For comparison, assuming Z = 0.008, then the bulk of stars in the $(i', 
r'-i')$ CMDs have ages log(t$_{yr}) < 8.5$, while stars 
that are older than log(t$_{yr}$) = 8.5 are resolved in the $(z', i'-z')$ CMDs.

\subsubsection{Comparing the disks of M81 and M82}

	The stars in the disks of M81 and M82 form a fossil record that can be mined to 
investigate the impact of the interaction between these galaxies, and a 
comparison of their CMDs can be used to identify differences in their star-forming 
histories in a purely empirical manner. In Figure 7 the 8 -- 
10 kpc CMDs of M81 and the CMDs of the 4 -- 6 kpc interval in M82 from Davidge (2008a) 
are compared with isochrones from Girardi et al. (2002; 2004). The 4 -- 6 kpc M82 observations 
cover 10 arcmin$^2$, whereas the 8 -- 10 kpc M81 observations cover 25 arcmin$^2$.
A distance modulus of 27.95 has been adopted for M82 (Sakai \& Madore 1999), as has 
a line of sight extinction A$_B = 0.12$ (Burstein \& Heiles 1982). 
The $(i', r'-i')$ and $(z', i'-z')$ CMDs of these galaxies come from the 
same MegaCam image, and so these data have the same image quality. 
Incompleteness sets in at roughly the same magnitude in the MegaCam CMDs of both galaxies, 
because the stellar density in each field is relatively low, and 
completeness is defined by photon statistics, rather than crowding. 
The near-infrared CMDs are constructed from data with the same 
instrument recorded during the same run. 

	The CMDs in Figure 7 indicate that the disks of M81 and M82 have experienced 
very different star-forming histories during the past $\sim 1$ Gyr. The upper 
envelope of points in the $(i', r'-i')$ and $(z', i'-z')$ CMDs of 
M81 indicates that a large population of stars with ages log(t$_{yr}$) $< 8.0$ is 
present. However, a corresponding population is clearly abscent in the M82 CMDs. Differences 
are also apparent when the AGB contents of the galaxies are compared, and 
this is evident from the $(M_K, J-K)$ CMDs of M81 and M82, which 
are compared in the bottom panel of Figure 7. While the majority of AGB stars in the 
disk of M81 have an age log(t$_{yr}) \geq$ 8.5, there is a clear concentration of 
objects in the M82 CMD that have log(t$_{yr})$ between 8.0 and 8.5.

	There is a modest spray of objects that extends to the right of the 
isochrones in the infrared CMDs, and these red objects hint at 
further differences between the stellar contents of the two disks.
The majority of stars with $J-K > 1.4 - 1.5$ in nearby galaxies tend to be C stars 
(e.g. Hughes \& Wood 1990; Demers, Dallaire, \& Battinelli 
2002; Davidge 2003; Batinnelli, Demers, \& Mannucci 2007). 
The isochrones do not sample this region of the $(K, J-K)$ CMD, as 
they are based on models that assume O-rich atmospheres, 
and so do not track the photometric properties of stars with C-rich atmospheres. 

	The region of the near-infrared CMD of the LMC discussed by Nikolaev \& Weinberg (2000) 
that contains a distinct C star sequence is indicated in the $(M_K, J-K)$ CMDs in Figure 7. 
The upper envelope of sources with $J-K > 1.5$ in the M82 CMDs tracks 
reasonably well the bright limit of the LMC C star box, and this supports 
the notion that the red stars in M82 are C stars. While incompleteness prevents an exploration 
of the full population of C stars in these galaxies, it is clear that 
only a modest number of objects fall within the C star box in the M81 CMD, while 
the corresponding portion of the M82 CMD is richly populated. A visual comparison of the 
$(M_K, J-K)$ CMDs in Figure 7 further indicates that the faint limit of the M81 data 
may extend to larger magnitudes than that of M82, and so the relative number of C stars in 
M82 with respect to those in M81 may even be greater than seen in Figure 7. 

	The differences between the C star contents of the disks of M81 and M82 are 
even more extreme than might be inferred from a simple visual comparison of their near-infrared 
CMDs, as a large fraction of the sources in the bright portion of the C star box in the M81 
8 -- 10 kpc $(M_K, J-K)$ CMD are foreground stars or background galaxies. To 
demonstrate this point the number of objects with M$_K$ between --8.5 and --9.0 and $J-K$ 
between 1.4 and 2.0 in the near-infrared CMDs of M81 and M82 were counted. These particular 
magnitude and color limits were adopted to sample an area of the CMD that contains C 
stars, while also avoiding regions where incompleteness occurs. 
Object counts were also made in a control area that is well removed from both galaxies to 
assess contamination from background galaxies and foreground stars. After subtracting the 
number counts from the control field, there remains only $8 \pm 8$ candidate bright 
C stars in the 8 -- 10 kpc region of M81, confirming that most -- if not all -- 
of the sources in the brightest portion of the C star box in the lower left hand corner 
of Figure 7 are either foreground stars or background galaxies. There is thus no 
statistically compelling evidence of a large C star population in 
the 8 -- 10 kpc interval of M81. In stark contrast, there are $221 
\pm 16$ candidate bright C stars in the 4 -- 6 kpc region of M82.

	The C star contents of galaxies are shaped by metallicity 
and star-forming history. If metallicity is too high then C is not 
dredged up during AGB evolution, and a C star is not formed. This 
metallicity sensitivity has been proposed to explain the tendency for lower C star fractions 
to occur in galaxies of progressively earlier morphological types (e.g. Battinelli \& Demers 
2005). However, the RGB sequences in the outer regions of M82 (Sakai \& Madore 1999) 
and M81 (Williams et al. 2008) indicate that old and intermediate-age stars in these 
galaxies have comparable metallicities. Therefore, the difference in 
bright C star numbers is suggestive of differences in the star formation history of the 
M81 and M82 disks during intermediate epochs, in the sense that the SFR in M82 was higher 
than that in M81.

	In summary, the comparisons in Figure 7 indicate that while the disk of M81 
has had a higher SFR during the past $\sim 100$ Myr than the disk of M82, the situation was 
likely very different prior to this. The richer population of the brightest AGB stars, 
including C stars, in M82 may be due to an elevated episode of star formation immediately 
following the interaction with M81, before the star burst activity settled into the central 
regions of the galaxy, where it is seen today. This being said, C star 
production occurs over an extended range of ages in simple 
stellar populations (e.g. Maraston 1998), and so some of the C star content in M82 may predate 
the interaction with M81. While the two fields considered 
here have different stellar densities and subtend different 
areas on the sky, these factors are not able to explain the difference 
in stellar contents see here.

\section{THE SPATIAL DISTRIBUTION OF DISK STARS}

	Early studies of the spatial distribution of stellar content often relied on 
integrated light measurements, and this remains the only option available for galaxies 
that are too distant to be resolved into stars. However, comprehensive surveys of the 
brightest stars have become practical for nearby galaxies with the recent deployment of 
large format imagers on telescopes that regularly deliver good image quality. 
The ability to assign ages using the standard clocks that can only be identified from 
CMDs, coupled with the potential to investigate the distribution of stars with a common 
evolutionary pedigree (e.g. RSGs), makes the use of resolved stars a more powerful 
means of probing the spatial distribution of stellar content than integrated light.

	The spatial distributions of the brightest MS stars and RSGs in M81 
with respect to the unresolved light, which is dominated by older stars, can be used to 
search for areas that experienced systematically elevated or depressed 
episodes of star formation during the past few tens of Myr. A caveat 
is that tidal forces may have distorted the structure of M81 at large radii. 
Depending on the geometry of the interaction, material may also 
have been pulled out of the disk plane, rendering as problematic 
efforts to investigate the de-projected distribution of stars. 
These effects can be suppressed by investigating trends from azimuthally-averaged data.

\subsection{Mapping the Brightest Blue Stars and RSGs in the Young Disk of M81}

\subsubsection{MS Stars and BSGs}

	Young MS stars and BSGs can be identified from their locations on 
CMDs. Following the criteria defined by Davidge (2008b) to investigate bright 
blue stars in the M81--M82 debris field, objects that have $r'-i'$ between --0.5 
and 0 and $i'$ between 24 and 22 are identified as massive MS stars 
and BSGs. Davidge (2008b) demonstrated that contamination from background galaxies 
in this part of the CMD is modest, making blue objects a powerful probe of the 
low-density outer regions of disks, where the contrast between objects that belong 
to a galaxy and those that do not is of prime importance.

	The distribution of bright blue stars in M81 is shown in Figures 8 (observed 
distribution) and 9 (de-projected distribution). 
The absence of bright MS stars at very small radii 
in Figures 8 and 9 is a consequence of the difficulties resolving even the brightest 
stars within the central $\sim 2$ kpc of M81 during natural seeing conditions. 
The algorithim used to construct the de-projected stellar distribution 
assumes that the disk is a plane with negligible thickness. Objects that are 
in parts of the disk that are warped out of the plane or that are in the tidal debris field, 
and hence may not be coplanar with the M81 disk, will not be correctly located in the 
de-projected image.

	Bright blue stars tend to fall along a ring in the inner regions of M81, 
and Davidge (2006a) identified the northern end of this feature as an area of 
long-term star-forming activity. The blue objects outside of this ring are concentrated along 
the spiral arms, with the highest density of bright blue stars in the north east spiral arm. 
The de-projected distribution of blue stars is consistent with 
morphological type Sbc (S. van den Bergh 2008, private 
communication), which is considerably later than the conventional classification of M81. 
Of course, there is a bias to a later morphological type if only the bluest, youngest 
objects are considered, and features such as the bulge and inter-disk light are neglected.

	Structures in the tidal debris field, including Ho IX, BK3N, the Arp Loop, M81 
West, and the Southern Tidal Arm are prominent features in the bright blue star 
maps, as might be expected given that they are sites of recent star formation 
(Davidge 2008b). The diffuse stellar groupings identified by Davidge (2008b) and 
Durrell et al. (2004) fall outside of the area shown in Figure 8. The modest density of 
foreground stars and background galaxies with blue colors is demonstrated in the 
small number of sources in the blue star maps outside of the main body of M81 and away from 
known tidal structures (Davidge 2008b). 

	Massive MS stars produce much of the UV emission in galaxy disks, and so 
it is not surprising that there is good agreement between 
the distribution of resolved blue stars and the GALEX 1516 \AA\ image of M81 
from Gil de Paz et al. (2007), which is also shown in Figure 8. Not only does 
the GALEX image show a ring of UV emission around the nucleus, but the relative 
strengths of the spiral arms in the UV are consistent with the observed density of 
MS stars. More specifically, the north east spiral arm is a source of stronger UV 
emission than the south west arm.

	Areas of recent star formation are sources of photons that heat 
dust to temperatures log(T$_{eff}) \sim 2 - 3$, and 
so a correlation between the distribution of MS stars and infrared 
emission might also be expected. The Spitzer 24$\mu$m 
and 160$\mu$m images of M81 from the Sings Fifth Data release are shown in Figure 8, and 
the correlation between the overall distributions of infrared light and bright blue stars 
is not as tight as that between UV emission and bright blue stars. The looser spatial 
correlation between blue stars and infrared emission is because sources other than young 
stars (e.g. AGB stars) can also heat dust to these temperatures. The northern end 
of the eastern spiral arm, which contains the densest concentration of MS stars in M81, 
is only moderately bright at $24\mu$m, while the strongest $24\mu$m emission tends to occur 
in both spiral arms along the major axis of the galaxy. 
Gordon et al. (2004) argue that the areas of most intense $24\mu$m emission may be regions 
of obscured star formation.

	Tidal structures such as the Arp Loop, Ho IX, 
M81 West, BK 3N, and the Southern Tidal Arm, which are clearly visible in the blue star 
map, are either weakly defined or not visible in the $24\mu$m image. However, the 160$\mu$m 
image shows that these are either sites of, or are close to areas of, emission 
from cool dust. The cool dust temperatures may indicate that dust has been 
displaced from the areas of active star formation. Simulations predict that 
tidal dwarfs likely do not have substantial dark matter 
halos (e.g. Barnes \& Hernquist 1992; Wetzstein, Naab, \& Burkert 2007), and 
so their gravitational potential wells should be shallower than those of 
galaxies that formed within dark matter halos. At a given SSFR then tidal dwarfs 
should be less able to retain interstellar material 
than dark matter-dominated systems. In addition, systems with shallow mass profiles may also 
only convert a minimal amount of gas into stars before stability of the 
interstellar component against star formation is restored (e.g. Kaufmann, Wheeler, 
\& Bullock 2007; Taylor \& Webster 2005), and so tidal dwarfs might also contain 
inherently large reservoirs of cool gas and dust until they are disrupted. 

\subsubsection{RSGs}

	RSGs form a prominent sequence in the CMDs of systems with ages $\geq 10$ Myr, 
and the bright portion of the RSG sequence forms a distinct finger 
that rises out of the complex of stars that dominates the faint end of the 
$(i', r'-i')$ and $(z', i'-z')$ CMDs (\S 3). Comparisons with isochrones in Figure 5 
suggest that this RSG finger is populated by stars with ages $\leq 100$ Myr. Fainter, 
and older, RSGs are mixed with the brightest AGB stars in the complex of objects near the 
faint limit of the CMDs. Given the broad range of ages covered by RSGs in the 
MegaCam data, it was decided to investigate the spatial distribution 
of RSGs in three separate magnitude and color intervals, with the goal of probing how the 
distribution of RSGs changes with age. The bright half of the RSG finger, which contains 
stars with approximate ages between 10 Myr and 30 Myr, is sampled by stars with $i'$ 
between 20.0 and 21.5, and $r'-i'$ between 0.6 and 1.0. The lower portion 
of the RSG finger, which contains stars with approximate ages between 30 and 100 Myr, 
is sampled with stars having $i'$ between 21.5 and 23.0, and $r'-i'$ between 0.3 and 0.8. 
Finally, the spatial distribution of the oldest RSGs and the brightest 
AGB stars is sampled with objects having $i$ between 23 and 24, and $r'-i'$ 
between 0 and 1.0. These objects have ages that exceed $\sim 100$ Myr.

	The observed and de-projected distributions of RSGs in M81 are shown in Figures 
8 and 9. For clarity, only RSGs with $i' > 23$ (i.e. those that are on the RSG 
finger above the mix of fainter RSGs and AGB stars) are shown in Figure 8. 
Contamination from background galaxies, which tend to have red colors in the magnitude 
range of interest, can be significant in the RSG samples, especially amongst the faintest 
RSGs. Indeed, a large number of objects, many of which are probably 
unresolved background galaxies, are seen outside of the M81 disk 
in the RSG maps in the lower row of Figure 9, in direct contrast with the modest number 
of objects in the outer regions of the blue star map.

	The distribution of RSGs in the top right hand corner of Figure 9 shows that 
-- like the brightest blue stars -- the brightest RSGs in M81 are good tracers of spiral 
structure. Like the brightest blue stars, a number of the brightest 
RSGs are also located in the young stellar ring in the inner disk. 
There is a distinct clumping of AGB stars and the faintest RSGs in the portion 
of the M81 disk that is closest to Ho IX, indicating that this was an area of active star 
formation at least 100 Myr in the past. Similar concentrations are also seen in 
the distribution of blue stars and brighter RSGs in the spiral arm that is closest to Ho IX. 
Given that the disk of M81 is rotating, and that the stellar concentration in the M81 disk 
that is presumably associated with Ho IX contains stars that span a range of ages, then it is 
likely that Ho IX is co-rotating with the M81 disk, which is consistent with it 
being of tidal origin.

	Differences between the spatial distributions of MS stars and RSGs become more 
apparent as fainter (and hence older) RSGs are considered, in the sense that spiral structure 
becomes more blurred as progressively older RSGs are mapped, due to the random velocities 
that are imparted to stars as they interact with gas clouds. It can also be seen 
in Figure 9 that the ratio of blue stars to fainter RSGs varies across the disk, in the 
sense that the number of blue stars with respect to RSGs decreases towards smaller radii 
in the inner regions of M81; this trend is quantified in \S 4.2.
Differences between the distribution of blue stars and RSGs are perhaps 
most obvious in tidal debris field objects. Whereas 
M81 West and the Southern Tidal HI arm appear as distinct concentrations of MS stars 
(\S 4.1.1), they contain only a modest number of RSGs. This is consistent with 
the notion that these structures are dominated by very young stars.

\subsection{The Radial Distribution of Blue Stars and RSGs}

\subsubsection{Star counts} 

	The radial behaviour of MS stars and RSGs can be investigated in a quantative manner 
using star counts that are azimuthally averaged over the entire disk. Such averaging allows 
global radial trends in stellar content to be examined while suppressing the impact of 
structures such as spiral arms. The influence of tidal features, which may 
be restricted to only a portion of the disk, on global radial trends are also reduced. 

	The luminosity functions (LFs) of blue stars and RSGs are shown in Figures 10 and 
11; the entries plotted in these figures are the number of stars per 0.5 magnitude 
interval per arcmin$^2$ in each annulus. A single range of colors, with $r'-i'$ between 
0.2 and 1.0, was used to generate the RSG LF. The LFs have been corrected for contamination 
from foreground stars and background galaxies by subtracting the LFs of objects 
in the blue star and RSG color intervals in a control field outside of the HI tidal 
debris field. After correcting for this contamination, statistically 
significant numbers of blue stars and RSGs are detected out to R$_{GC} \sim 18$ kpc. 
The young stellar disk of M81 thus extends to $\sim 10$ disk scalelengths, which is comparable 
to the NGC 300 disk (Bland-Hawthorn et al. 2005).

	The number densities of blue stars and RSGs 
do not change substantially throughout much of the central 10 kpc of M81. This is demonstrated 
in Figure 12, where the radial behaviour of the mean number of sources with $i'$ 
between 22.25 and 23.75 in Figure 10 and $i'$ between 21.25 and 22.75 in 
Figure 11 are examined. The differences between the radial distribution of blue stars and RSGs 
are highlighted in the lower panel of Figure 12, where the ratio of blue to red star counts is 
shown. This ratio is constant to within $\pm 0.1$ dex between R$_{GC} =$ 4 and 12 
kpc. However, the ratio of blue stars to RSGs drops by 0.5 dex when R$_{GC} < 4$ kpc, and 
there is also a steady decline in this ratio with increasing radius when R$_{GC} > 12$ kpc. 
M81 is not the only galaxy to show such trends, as previous studies of the brightest resolved 
stars in galaxies have found that stars in galaxy disks with different ages may have 
different radial distributions (e.g. Davidge 2006b; de Jong et al. 2007). 
The differences in the number densities of MS stars and RSGs in Figure 12 suggest that the 
SSFR across the disk of M81 was not uniform during 
the past few tens of Myr, but varied with radius such that the 
brightest blue stars are deficient with respect to RSGs at both small and large radii. 

\subsubsection{Specific frequency measurements and evidence for an age gradient}

	The mix of young and old stars throughout the M81 disk can be investigated by 
computing the number of blue stars and RSGs per unit integrated total brightness in 
each annulus, which is a measure of their specific frequency (SF). For this study, the 
SF measurements use the $K-$band surface brightness profile of M81 
from Jarrett et al. (2003), and are normalized to M$_K = -16$. Light in 
the $K-$band is dominated by stars that formed during intermediate and early epochs (e.g. 
Maraston 1998). Therefore, systematic radial variations in the SFs of blue stars and RSGs 
measured with respect to $K-$band light are a signature of population gradients due to 
departures of the recent SFR from the mean SFR measured over Gyr or longer timescales.

	The Jarrett et al. (2003) data trace the $K-$band 
light of M81 out to 750 arcsec (i.e. R$_{GC} = 13$ kpc). $K-$band surface 
brightnesses in the radial interval 13 -- 15 kpc were computed from the 
$H-$band profile by adopting the $H-K$ color of M81 at smaller radii. 
The $H-$band profile was extrapolated to compute surface 
brightnesses in the 15 -- 18 kpc interval. 
The SF measurements are thus very uncertain when R$_{GC} > 15$ kpc.

	The SFs of blue stars and RSGs are compared in Figure 13. There is a tendency 
for the SFs of both stellar types to increase with increasing R$_{GC}$, indicating that 
the ratio of young to old$+$intermediate-age stars 
increases with increasing radius. This trend starts at relatively small radii, and so is 
not a manifestation of the extrapolation of the near-infrared surface brightness profile 
at large radii. Thus, the SF measurements further confirm that young stars are not uniformly 
mixed throughout the M81 disk, but tend to appear in larger numbers as radius increases.

	The SFs of blue stars and RSGs climb dramatically when R$_{GC} > 14$ 
kpc, which is where the SF measurements are most uncertain. Still, 
there is evidence from both the star counts and integrated light photometry that the 
radial stellar distribution in M81 changes near R$_{GC} \sim 13 - 14$ kpc. In particular, 
the $J$ and $H$ profiles of M81 in the 2MASS Large Galaxy 
Atlas are similar out to 800 arcsec ($\sim 14$ kpc), with 
both showing a downward break in surface brightness at 750 arcsec ($\sim 13$ kpc). 
The relations between the mean densities of MS stars and RSGs with radius in Figure 12 
also both show a kink near 750 arcsec.

	The SF trend at large radius can be checked with deeper photometric measurements. 
RGB stars are among the brightest members of the populations that dominate 
the integrated $K-$band light, and are tracers of old/intermediate age stars. 
The SF curves in Figure 13 predict that the ratio of blue stars and/or RSGs to 
RGB stars should increase with progressively larger R$_{GC}$ in M81, 
and that the trend will steepen when R$_{GC} > 14$ kpc.
Deep wide-field images in which RGB stars can be resolved will thus allow the behaviour 
of the SF curves at large radii in Figure 13 to be checked in an unambiguous manner. 

\subsubsection{Differences in the SSFRs of M81 and NGC 2403}

	SF measurements can be used to compare the SSFRs 
of galaxies in an empirical manner. This is demonstrated in Figure 14 where the SFs of 
blue stars and RSGs located between R$_{GC} = 4$ and 10 kpc in M81 are compared with the 
SFs of similar stars located between R$_{GC} = 6$ and 12 kpc in the Sc galaxy NGC 2403. 
The NGC 2403 SFs are taken from the MegaCam observations discussed by Davidge (2007), and 
the magnitudes of stars in NGC 2403 have been shifted to account for the different 
distance modulus of M81.

	The SF curves of both galaxies are power laws that have similar exponents. 
However, the SF measurements of M81 consistently fall $\sim 1$ dex below those of NGC 2403. 
To the extent that both galaxies have similar underlying stellar 
contents that contribute the bulk of the $K-$band 
light, then this offset indicates that the SSFR in M81 
during the past $\sim 30$ Myr was a factor of ten 
lower than in NGC 2403. If this is representative of the mean difference in 
SSFRs between these galaxies throughout the age of the Universe then M81 is dominated by 
an older population of stars than NGC 2403.

	The difference in the SF curves is consistent with 
the relative $24\mu$m fluxes of these galaxies, which gauges the recent SFR, and their 
total $K-$band brightnesses, which measures total stellar mass. 
The $24\mu$m fluxes of M81 and NGC 2403 are comparable (Gil de Paz et al. 2007), 
indicating that they have comparable {\it total} SFRs. However, the 
integrated $K$ brightnesses of M81 and NGC 2403 differ by 2.7 
magnitudes, in the sense that M81 is brighter (Jarrett et al. 2003), indicating 
that their total stellar masses differ by roughly a factor of ten. 
Thus, the ratio of $24\mu$m flux per unit mass in NGC 2403 is an order of magnitude higher 
than in M81, in agreement with the difference in SFs shown in Figure 14.

	The SFs of RSGs and bright MS stars in M81 and NGC 2403 indicate that, 
despite the cosmologically-recent interaction with M82, the 
present-day SFR of M81 is not abnormally high. 
Brinchmann et al. (2004) and Asari et al. (2007) investigate the relation 
between SFR and galaxy mass, and their results suggest that the
SSFRs of galaxies with masses that differ by 1 dex typically differ by a few tenths to 1 dex, 
with the large dispersion reflecting the wide range of star-forming histories 
seen among Sb and later galaxies (e.g. Figure 6 of Kennicutt, Tamblyn, and Congdon 1994). 
Thus, the SSFR of M81 is comparable to that of a field galaxy. 

	Woods \& Geller (2007) investigate the impact of 
interactions on galaxy pairs, and find that while the SSFRs in galaxies in 
pairs are higher than of those in the field, the enhancement in the SSFR is small
for galaxies with M$_z \leq -22$. Given that M$_{z'} < -22$ for M81, then 
a substantial increase in the SSFR of M81 with respect to non-interacting galaxies 
would not be expected. It has also been $\sim 0.2$ Gyr since M81 interacted 
with M82, and this is twice the typical timescale for sustained starburst activity 
(e.g. Marcillac et al. 2006). 

\section{DISCUSSION \& SUMMARY}

	Data obtained with the CFHT MegaCam and WIRCam imagers have been used to 
survey the brightest stars in the disk of M81. With the exception of the central 
$\sim 2$ kpc of the galaxy, where crowding prevents the detection of 
individual stars, and the gaps between CCDs, the MegaCam 
data are spatially complete to a radial distance of 18 kpc; at larger 
radii the southern portions of M81 are outside of the MegaCam 
field. The WIRCam images cover roughly one third of the M81 disk 
out to R$_{GC} \sim 18$ kpc. The key results of this study are discussed and summarized below.

\subsection{The Outer Disk of M81}

	Young stars have been traced out to R$_{GC} \sim 18$ kpc in M81, and 
the density distribution of these objects does not follow a single 
exponential profile. This is reminiscent of what is seen in 
the integrated light profiles of disks, which show a diverse array of behaviours at large radii. 
In the majority of cases the exponential light profile that is the structural 
hallmark of disks changes slope at large radii, typically 
breaking downwards (i.e. the exponential decline steepens), although 
the light profile breaks upwards (i.e. the exponential decline 
becomes shallower) in roughly a third of spiral galaxies (e.g. Pohlen \& Trujillo 2006). 

	The near-infrared light profiles of M81 constructed from data in the 
2MASS Large Galaxy Atlas steepen near 14 kpc, and this break radius 
is near the high end for galaxies in general (Pohlen \& Trujillo 2006). 
The tendency for the light profile to steepen, rather than flatten, at large radius 
is not common among Sb galaxies (Pohlen \& Trujillo 2006). In contrast, the rate of decline of 
RSGs and bright MS stars flattens near R$_{GC} \sim 14$ kpc, 
indicating that the radial distribution of old stars, which dominate the near-infrared 
light, and young stars in M81 differ, in that the latter have a flatter 
distribution than the former. Young stars thus account for a progressively larger 
fraction of the total stellar content in the outer disk of M81 as R$_{GC}$ increases. 
Kong et al. (2000) and Jiu-Li et al. (2004) investigate the stellar 
content of M81 using images recorded through narrow-band 
filters, and also conclude that mean age decreases with increasing R$_{GC}$ in M81. 

	Age gradients, in the sense of younger mean ages towards larger radii, are common in the 
disks of spiral galaxies (e.g. Bell \& de Jong 2000). The SF measurements in \S 4 suggest 
that the outermost regions of the M81 disk may contain a particularly rich population of 
young stars, as the contribution made by young stars increases significantly near 14 kpc. 
A preponderance of resolved young stars with respect to older populations 
at large radii is also seen in the disks of NGC 247 (Davidge 2006) and NGC 4244 (de Jong et 
al. 2007). The disk of NGC 4244 is of particular interest as it is viewed edge-on, so that 
height above the disk plane provides another dimension for structure studies that can 
be used to shed light on the nature of disk truncation. 
de Jong et al. (2007) find that the radius at which star counts in the disk 
NGC 4244 drop off is not a function of stellar age or height above the disk plane, 
and argue that disk truncation in this galaxy is probably due to recent 
or on-going dynamical processes that must affect the spatial distribution 
of stars in both the thin and thick disks. In the particular case of M81, the spatial 
distributions of the young and old$+$intermediate age populations indicate that their 
angular momentum distributions differ, such that the young star distribution is 
skewed to a higher mean value than that of old$+$intermediate age 
stars. One rationalization of this is that torques 
exerted by M82 could distort the angular momentum distribution of the interstellar 
medium of M81, and this legacy would be passed down to the kinematic properties of stars 
that subsequently form from this material. Younger at el. (2007) discuss the formation 
of extended exponential disks during minor mergers. Their simulations predict that 
stars move outward in the disk as gas is funneled inward. In the absence of external gas sources 
this would result in older mean ages towards larger radii, which is contrary 
to what is seen in M81.

	The local SSFR measured from UV images appears to increase with 
radius in many nearby galaxies (e.g. Munoz-Mateos et al. 2007).
Still, the flatter distribution of young stars with respect to the older stellar body 
found in studies of resolved stars may be in conflict with the results from visible 
integrated light studies. Bakos, Trujillo, \& Pohlen (2008) find that the color profiles 
of disks usually reverse at the point where the light profile changes, in the sense that 
$g'-r'$ increases towards larger radii. While it it difficult to compute reliable 
integrated colors at very large radii in M81, the change 
in the SF of RSGs and MS stars at large radii in Figure 13 
suggests that the integrated $g'-r'$ color of M81 probably becomes 
progressively bluer (i.e. smaller $g'-r'$) with radius when R$_{GC} > 14$ kpc.

\subsection{The Metallicity of Young Stars in M81}

	Secular processes and galaxy-galaxy interactions re-distribute 
material in galaxy disks, and thereby alter gradients that may have been imprinted early-on. 
The tidal HI features near M81 are consistent with 
the large-scale redistribution of gas in this galaxy. 
If gas in the M81 disk was stirred by tidal interactions then this would 
flatten any metallicity gradients that were in place prior to this event. In fact, 
the color of the RSG plume in M81 does not change when R$_{GC} > 4$ kpc, suggesting 
that the metallicity of the material from which RSGs formed may not vary with radius, 
remaining near one half solar. The absence of a significant abundance gradient 
is consistent with previous studies, which find at most only minor abundance gradients 
when R$_{GC} > 4$ kpc. While various studies have found 
that [O/H] measured in HII regions tends to drop with increasing 
radius in M81 (e.g. Stauffer \& Bothun 1984; Garnett 1986; Zaritsky, Kennicutt, \& 
Huchra 1994), the relation between [O/H] and radius does not follow a simple 
power-law. Rather, while there is a pronounced radial [O/H] gradient in the inner 
disk, Figure 8 of Zaritsky et al. (1994) indicates that 
[O/H] is constant in the disk outside of one half the isophotal radius.
Using images recorded through narrow-band filters, one of which is
centered on the near-infrared Ca II triplet, Kong et al. (2000) find that the mean stellar 
metallicity in M81 does not change with radius, remaining fixed near 
a luminosity-weighted metallicity that is solar or higher.
Re-examining the same data, Jiu-Li et al. (2004) find that there may be only a very mild 
metallicity gradient, in the sense that the mean metallicity drops by 
only 0.1 dex from the bulge (Z = 0.022) to the outer disk (Z = 0.016). A caveat is 
that the metallicity measured from such integrated light studies is a luminosity-weighted 
mean that is dominated by stars that are markedly older than RSGs. 

	The Local Group galaxy M31 has a morphology and mass that is similar to M81, 
and likely experienced an interaction during intermediate 
epochs. It is thus also of interest to consider the radial distribution of metals in the 
disk of that galaxy. The relation between [O/H] and radius in the disk of 
M31 shows substantial scatter, with no obvious trend defined by HII regions at 
intermediate radii (e.g. Figure 8 of Zaritsky et al. 1994).
Abundance measurements from the spectra of F and B supergiants are 
consistent with no metallicity gradient (Venn et al. 2000; Trundle et al. 
2002). Moreover, while not sampling the inner disk of M31, the fields studied by 
Bellazzini et al. (2003) also suggest that the mean metallicity of 
RGB stars in the outer disk of that galaxy does not change with radius. These data hint that -- 
as in M81 -- interstellar material in M31 may have been tidally stirred.

\acknowledgments{Sincere thanks are extended to the anonymous referee for providing a quick 
and thorough report that greatly improved the manuscript.}

\appendix

\section{Reconciling a Post-Interaction Starburst in M81 with an Old Integrated Light Age Estimate}

	The CMDs of the 2 -- 4 kpc interval indicate that 
the present-day SSFR in this portion of M81 is relatively 
low when compared with the disk at larger radii. This may seem surprising given that 
star formation in interacting galaxies tends to be centrally concentrated (e.g. Iono, 
Yun, \& Mihos 2004; Kewley, Geller, \& Barton 2006).
In fact, evidence that there was a central burst of star formation comes in the form of 
the population of x-ray binaries discussed by Swartz et al. (2003). Still, 
the mass of stars that formed during intermediate epochs was not large enough to affect 
greatly the present-day integrated visible photometric properties of the central 
regions of M81. Indeed, the `old' age measured by Kong et al. 
(2000) and Jiu-Li et al. (2004) in the central few kpc of M81 indicates that the stars that 
formed a few hundred Myr in the past can not account for more than a few percent 
of the total stellar mass (e.g. Serra \& Trager 2007). 

	The evidence of elevated star-forming activity in the 
inner regions of M81 and the old age deduced from narrow-band 
photometry are not necessarily inconsistent, as a star-forming episode that involved 
a significant fraction of the gas in M81 would not contribute significantly to the mass 
in the central few kpc of the galaxy. The combined mass of HI and H$_2$ in a galaxy 
with a luminosity comparable to that of M81 is $\sim 10^{10}$ M$_{\odot}$
(Boselli, Lequeux, \& Gavazzi (2002). Adopting a star formation efficiency of 
$\sim 5\%$ (Inoue, Hirashita, \& Kamaya 2000), then 
$4 \times 10^8$M$_{\odot}$ would form from $10^{10}$ M$_{\odot}$ of gas in a single 
star-forming event. The integrated brightness of M81 in the central 2 kpc is $K = 4.6$, so 
that M$_K = -23.2$. Assuming $M/L_K = 1$, then the total stellar 
mass in this part of the galaxy is $4 \times 10^{10}$M$_{\odot}$. The mass of stars that 
would form if all of the ISM were channeled into the central few kpc of M81 would then only 
amount to $\sim 1\%$ of the original stellar mass, and so would not have a large impact 
on the integrated photometric properties a few hundred Myr later. Even if there were two or 
three distinct episodes of star formation of this nature then the total stellar mass 
formed would be only a few percent of that initially present.

\section{Infrared Photometry of M81 Globular Clusters}

	As the nearest large spiral galaxy outside of the Local Group, M81 is a prime 
laboratory for investigating the globular cluster system of a large early-type 
disk galaxy. While numerous cluster candidates have been identified from photometric 
data (e.g. Perelmuter \& Racine 1995; Davidge \& Courteau 1999), only a modest fraction 
of these have been confirmed spectroscopically as globular clusters. Even fewer 
of these have the near-infrared photometric measurements that are 
required to more fully characterize their spectral energy-distributions. 
The impact of interstellar extinction, which will be significant for clusters 
that are viewed through dust in disks, is also lower in 
the near-infrared than at visible wavelengths. Finally, from a more pragmatic perspective, 
many of the globular clusters that will be studied with the next generation 
of large ground-based telescopes in more distant galaxies will be investigated in 
greatest detail at near-infrared wavelengths, where adaptive optics (AO) systems will deliver 
near diffraction-limited image quality. It is thus important to 
characterize globular clusters in nearby galaxies at near-infrared wavelengths 
to provide a benchmark for studies of more distant cluster systems.

	The majority of spectroscopically confirmed globular clusters in M81 
are to the north and west of the galaxy center (e.g. Figure 1 of 
Schroder et al. 2002), and there are only five confirmed clusters in 
the WIRCam field. All of these are viewed against the main body of the stellar 
disk, and so extinction might be significant for some. 
The properties of the clusters in the WIRCam field with spectra discussed by Perelmuter, 
Brodie, \& Huchra (2005) and Schroder et al. (2002) are listed in Table 1. 
The source of the $V$ and [Fe/H] measurements are given 
in the last column. The near-infrared brightnesses and colors of 
clusters \# 50401 and 50415 were also measured by 
Davidge (2006a). The photometric properties of the CFHTIR and WIRCam measurements 
agree to within 0.01 -- 0.02 mag for cluster 50401, and to within 0.1 mag for cluster 
50415, indicating that there is reasonable photometric consistency.

	The clusters in Table 1 are not representative of the 
entire M81 globular cluster system. If M31 were viewed at the 
same distance as M81, then the $K-$band measurements discussed 
by Barmby, Huchra, \& Brodie (2001) indicate that its GCLF would peak near $K = 17.5$, with 
the majority of clusters having $K$ between 16 and 18.5. The clusters in 
Table 2 of Davidge (2006a) and Table 1 of the present study are in the bright tail 
of the GCLF. This is not unexpected, given the observational bias against 
spectroscopic studies of the faint members of the M81 cluster system. In addition, the 
majority of M31 clusters have $V-K$ between 2.0 and 2.5 and $J-K$ between 0.55 and 0.80
(Barmby et al. 2000). For comparison, 3 of the 
seven clusters in M81 that have IR photometry have $V-K > 2.5$. This is almost certainly 
not due to fundamental differences in the cluster system properties; rather, it is 
probably due to dust extinction in the disk of M81.

\parindent=0.0cm

\clearpage

\begin{table*}
\begin{center}
\begin{tabular}{cccccccc}
\tableline\tableline
Cluster \# & $K$ & $H-K$ & $J-K$ & $V$ & $V-K$ & [Fe/H] & Reference \\
\hline
50255 & 16.101 & 0.098 & 0.604 & 18.43 & 2.33 & --0.04 & P1995 \\
50401 & 17.111 & 0.161 & 0.751 & 19.93 & 2.82 & --0.04 & P1995 \\
50415 & 17.116 & 0.090 & 0.658 & 19.24 & 2.12 & --1.90 & P1995 \\
50418 & 16.287 & 0.194 & 0.887 & 18.45 & 2.16 & --1.09 & S2002 \\
50787 & 16.343 & 0.115 & 0.774 & 19.12 & 2.78 & --1.06 & S2002 \\
\tableline
\end{tabular}
\caption{Near-Infrared Photometry of Spectroscopically Confirmed Globular Clusters}
\end{center}
\end{table*}

\clearpage
\begin{figure}
\figurenum{1}
\epsscale{0.85}
\plotone{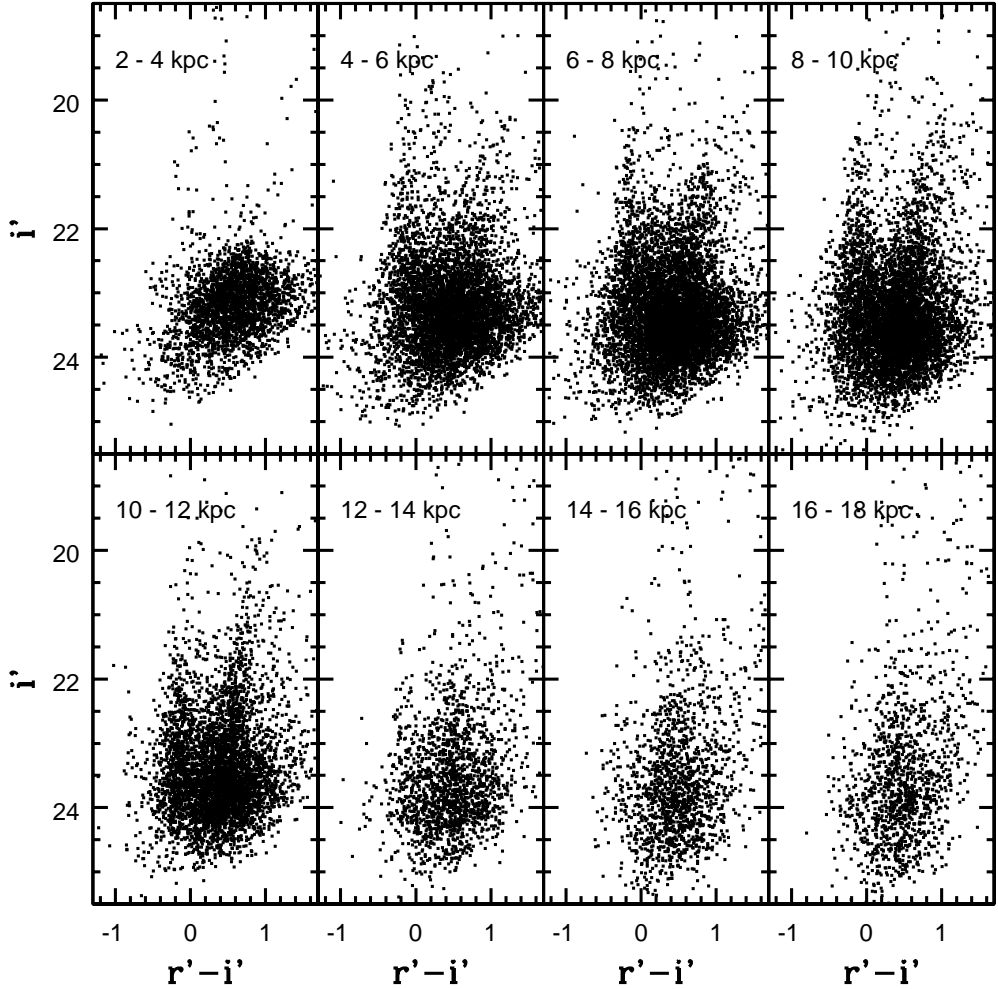}
\caption
{The $(i', r'-i')$ CMDs of sources in various annuli. Distances are 
in the M81 disk plane assuming an inclination of 59.3 degrees and a 
distance modulus of 27.8. Note the modest number of bright MS stars and RSGs 
in the 2 -- 4 kpc interval when compared with larger radii. Stars belonging to M81 
dominate the CMDs out to R$_{GC} \sim 14$ kpc. 
The CMD of the 16 -- 18 kpc interval is dominated by foreground stars 
and background galaxies, although objects belonging to M81 are still present (\S 4).}
\end{figure}

\clearpage
\begin{figure}
\figurenum{2}
\epsscale{0.85}
\plotone{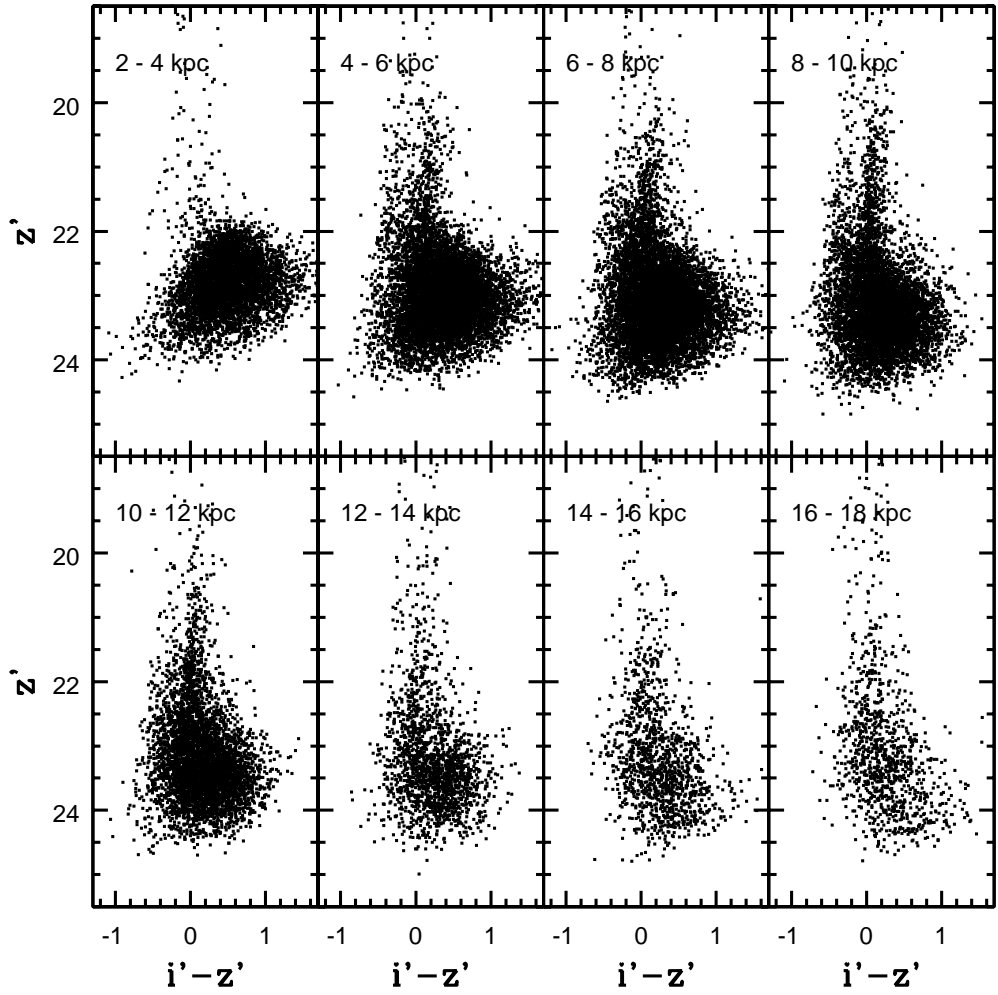}
\caption
{The same as Figure 1, but showing $(z', i'-z')$ CMDs.}
\end{figure}

\clearpage
\begin{figure}
\figurenum{3}
\epsscale{0.85}
\plotone{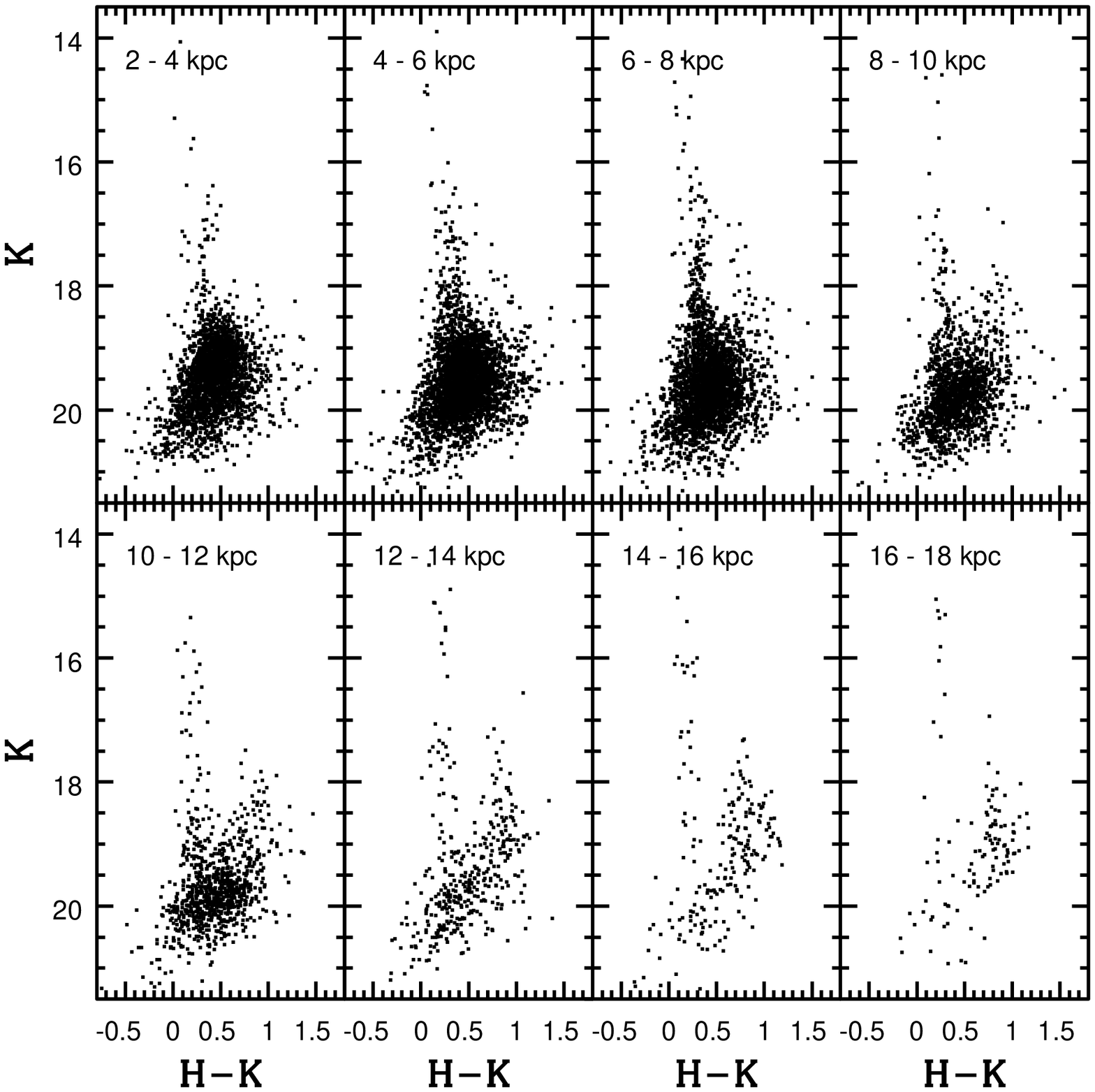}
\caption
{As in Figure 1, but showing $(K, H-K)$ CMDs. The WIRCam data cover roughly 
one third of the M81 disk.}
\end{figure}

\clearpage
\begin{figure}
\figurenum{4}
\epsscale{0.85}
\plotone{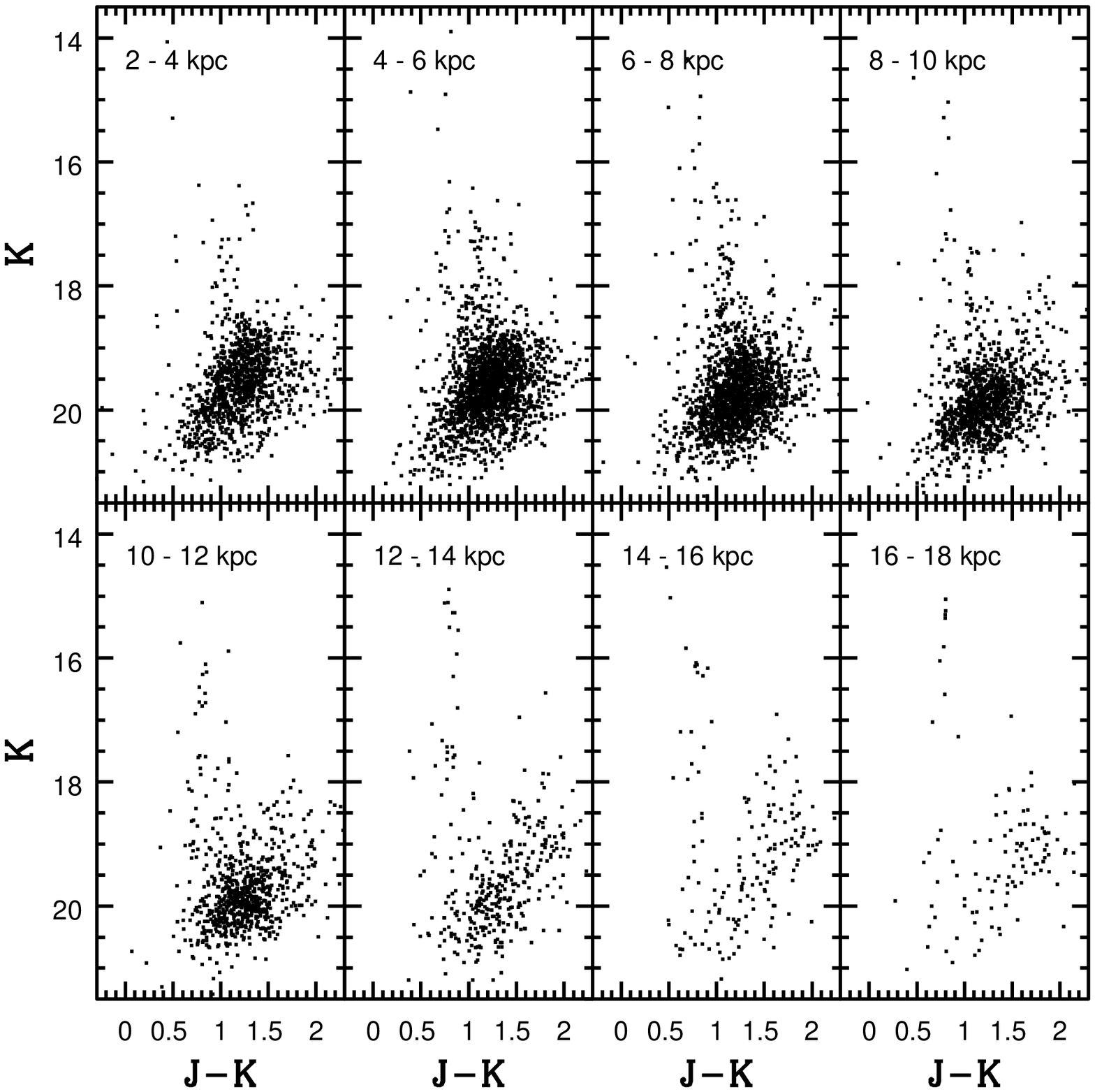}
\caption
{As in Figure 1, but showing $(K, J-K)$ CMDs. The WIRCam data cover roughly 
one third of the M81 disk.}
\end{figure}

\clearpage
\begin{figure}
\figurenum{5}
\epsscale{0.85}
\plotone{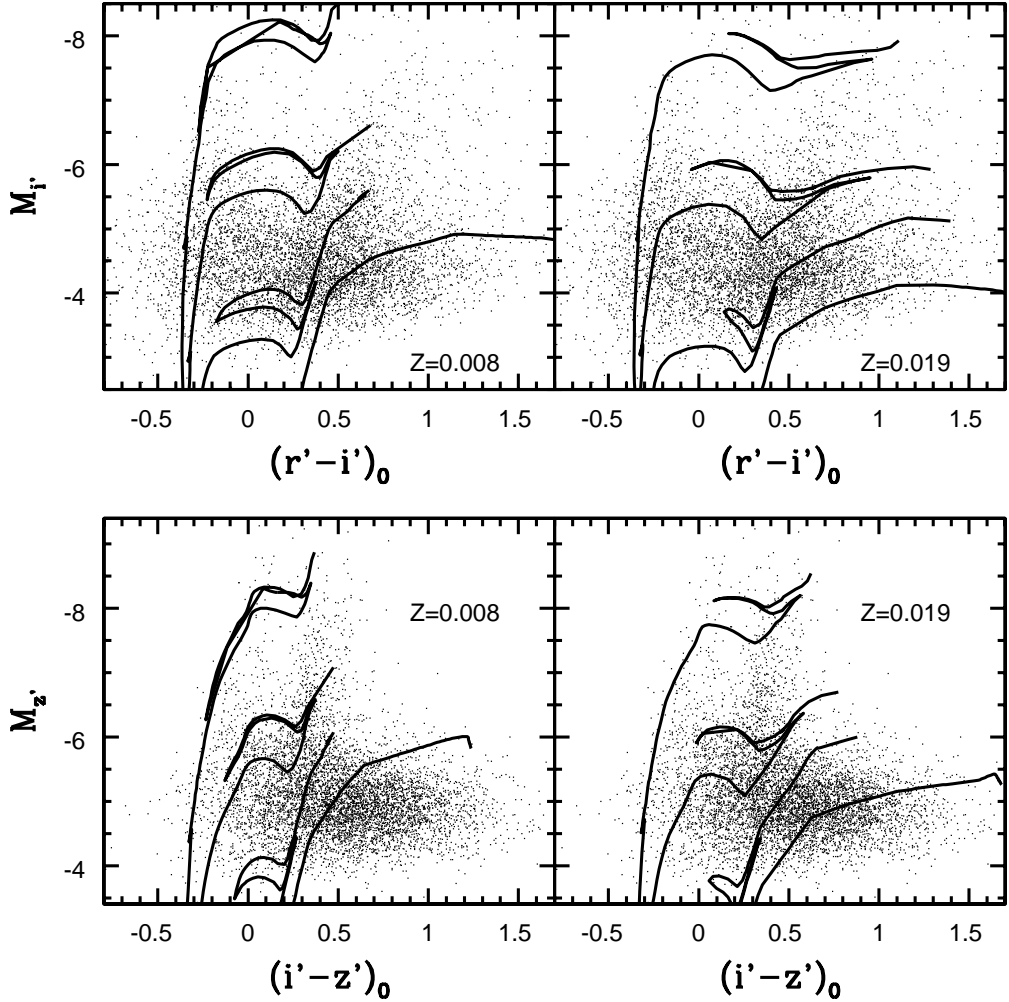}
\caption
{Isochrones from Girardi et al. (2004) with log(t$_{yr}) = 7.0, 7.5, 8.0,$ and 8.5 
and metallicities Z = 0.008 and Z = 0.019 are compared with the $(i', r'-i')$ and 
$(z', i'-z')$ CMDs of sources with R$_{GC}$ between 8 and 10 kpc. 
The blue envelope of points is matched by the log(t$_{yr}$) = 7.0 
sequences, while the color and curvature of the RSG sequence with M$_{i'} < -5$ and M$_{z'} 
< -5.5$ are matched by the Z = 0.008 models. This is in reasonable agreement with the 
metallicity of young stars calculated by Williams et al. (2008) in the outer disk of M81.} 
\end{figure}

\clearpage
\begin{figure}
\figurenum{6}
\epsscale{0.85}
\plotone{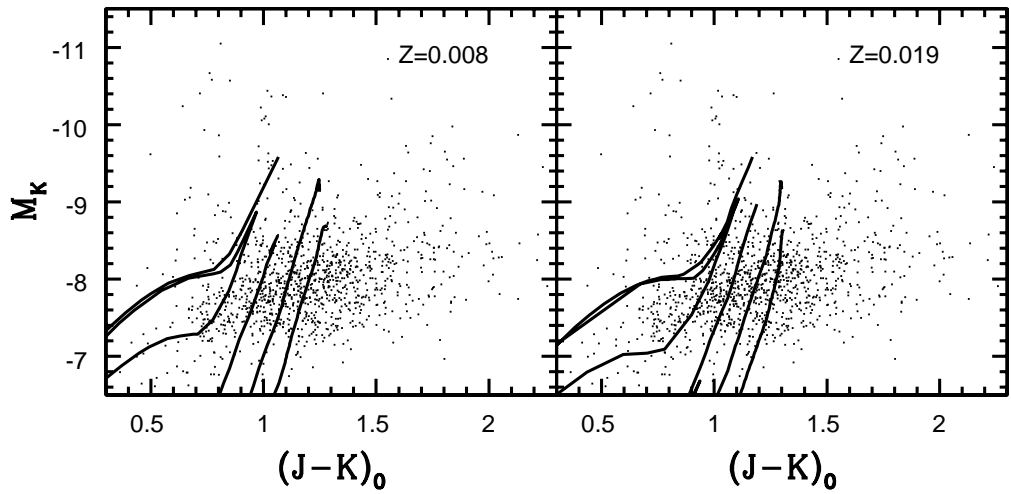}
\caption
{Isochrones from Girardi et al. (2002) with log(t$_{yr}) = 7.5, 8.0, 8.5,$ and 9.0
and metallicities Z = 0.008 and Z = 0.019 are compared 
with the $(K, J-K)$ CMDs of sources with R$_{GC}$ between 8 and 10 kpc. 
Note that AGB stars with ages log(t$_{yr}$) $\geq 9.0$ are present in these data.}
\end{figure}

\clearpage
\begin{figure}
\figurenum{7}
\epsscale{0.85}
\plotone{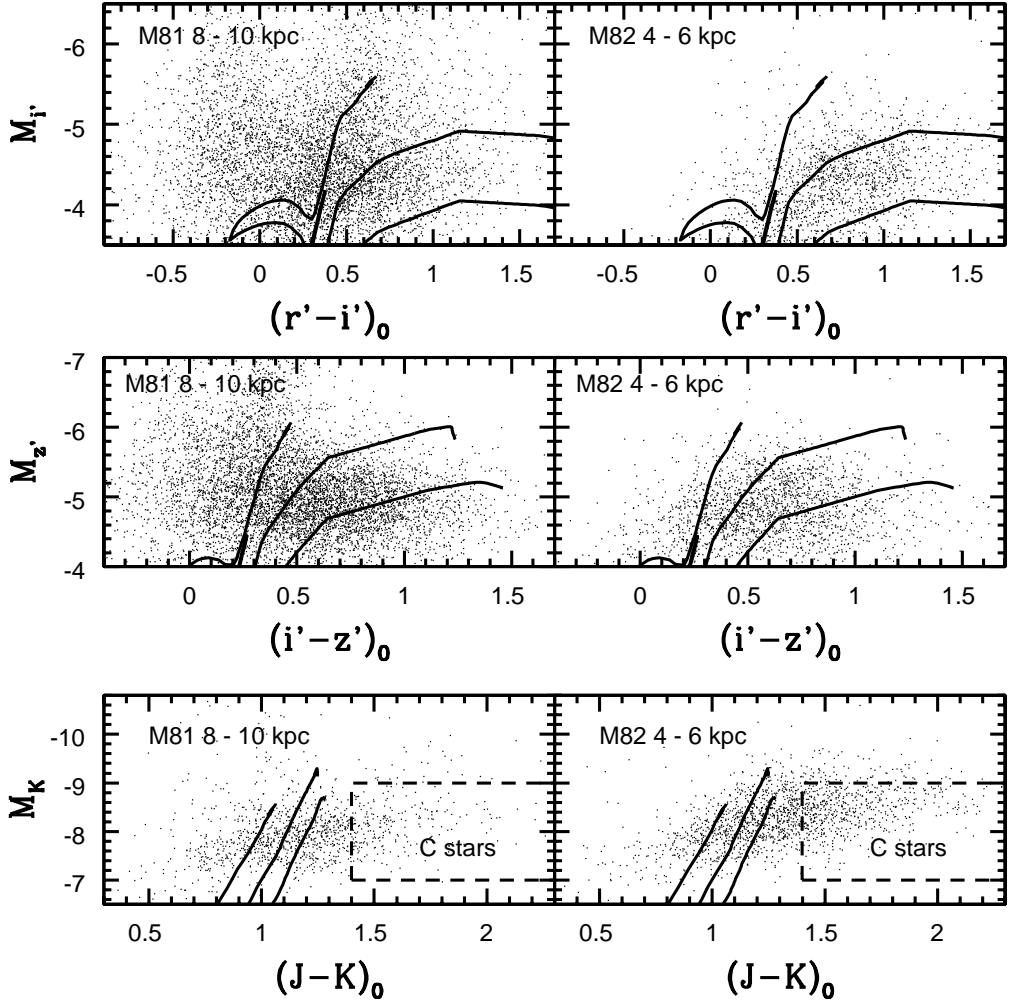}
\caption
{The CMDs of the 8 -- 10 kpc interval in M81 are compared with those of the 4 -- 6 kpc 
interval of M82 from Davidge (2008a). The isochrones are Z = 
0.008 sequences from Girardi et al. (2002; 2004) with log(t$_{yr}) = 7.5$, 8.0, and 8.5. 
The region of the $(M_K, J-K)$ CMD of the LMC from Nikolaev \& Weinberg (2000) 
that contains a clear C star sequence is enclosed by the dashed box on the $(M_K, J-K)$ CMDs.}
\end{figure}

\clearpage
\begin{figure}
\figurenum{8}
\epsscale{0.85}
\plotone{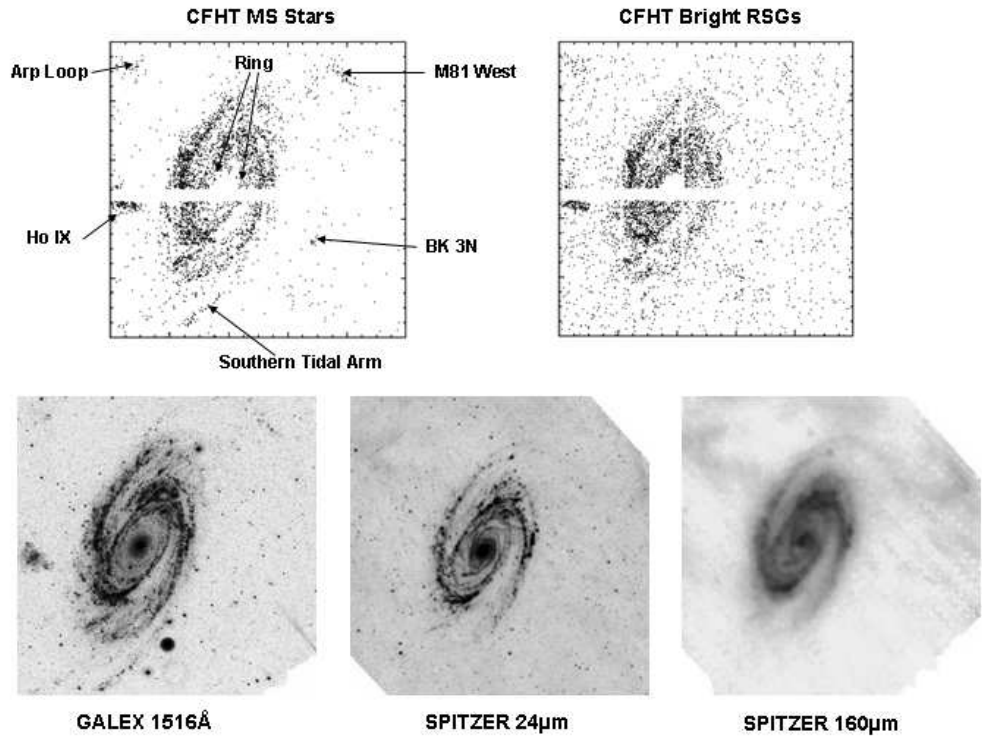}
\caption
{The distribution of MS stars and RSGs with $i' < 23$ in a $20 
\times 20$ arcmin section of the CFHT MegaCam field is shown in the upper row of the figure. 
The absence of bright MS stars and RSGs in the innermost regions of M81 is 
due to source confusion. The three diffuse stellar 
groupings identified by Davidge (2008a) are outside of the field shown here. 
UV and IR images of M81 from Gil de Paz et al. (2007) and the SINGS Fifth Enhanced 
Data Release are shown in the lower row. Prominent structures in the tidal 
debris field are identified, as are the eastern and western boundaries of the ring 
of young MS stars in the inner disk, which is a prominent feature in the 
GALEX image. The UV and IR images have the same angular scale as the MS and RSG maps.}
\end{figure}

\clearpage
\begin{figure}
\figurenum{9}
\epsscale{0.85}
\plotone{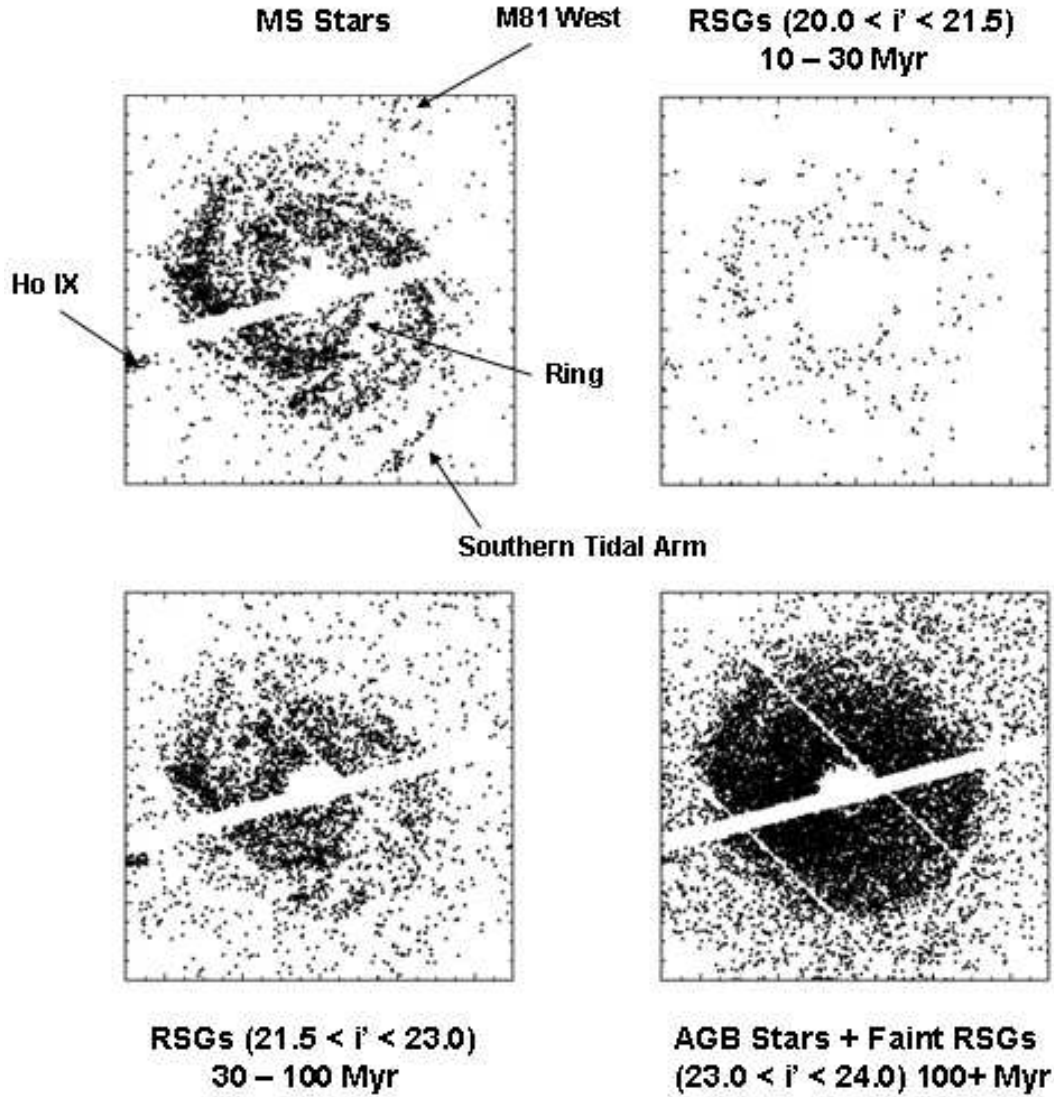}
\caption
{The de-projected distribution of MS stars and RSGs. The images have been rotated 
so that the major axis of M81 is on the vertical axis. Various structures in 
the M81--M82 debris field are identified. The spatial distributions of RSGs in three 
different brightness ranges are shown; the corresponding approximate ages of stars in each 
sample are also indicated. Note that blue stars and the youngest RSGs define a 
ring around the central regions of the galaxy, and tend to be located along the 
spiral arms, with the highest concentration of blue stars in the northern arm. 
RSGs with $i'$ between 21.5 and 23.0 also track spiral structure. There is a 
concentration of objects in the RSG $+$ AGB star map in the part of the disk that 
is closest to Ho IX, indicating that this was an area of concentrated star formation 
some $\sim 100$ Myr or more in the past. Finally, while M81 West 
and the Southern Tidal Arm are both prominent structures 
in the MS map, they are not as well-defined in the RSG maps.}
\end{figure}

\clearpage
\begin{figure}
\figurenum{10}
\epsscale{0.85}
\plotone{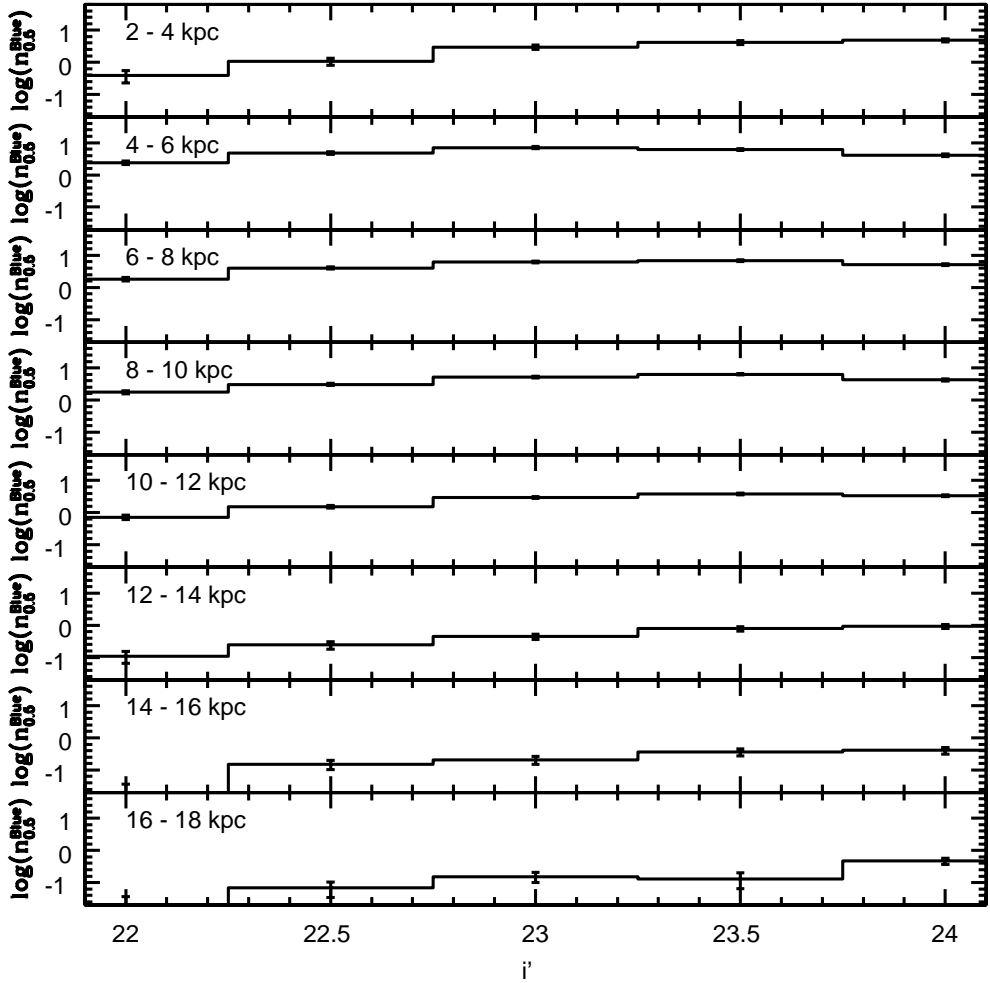}
\caption
{The LFs of stars that have $r'-i'$ between --0.5 and 0.0. 
n$_{0.5}^{Blue}$ is the number of objects per 0.5 mag $i'$ interval 
per arcmin$^{-2}$. The LFs have been corrected for contamination from foreground 
stars and background galaxies by subtracting the LFs of sources in a control field. 
The errorbars show Poisson uncertainties. Note that 
statistically significant numbers of objects are seen out to at least R$_{GC} = 18$ kpc.}
\end{figure}

\clearpage
\begin{figure}
\figurenum{11}
\epsscale{0.85}
\plotone{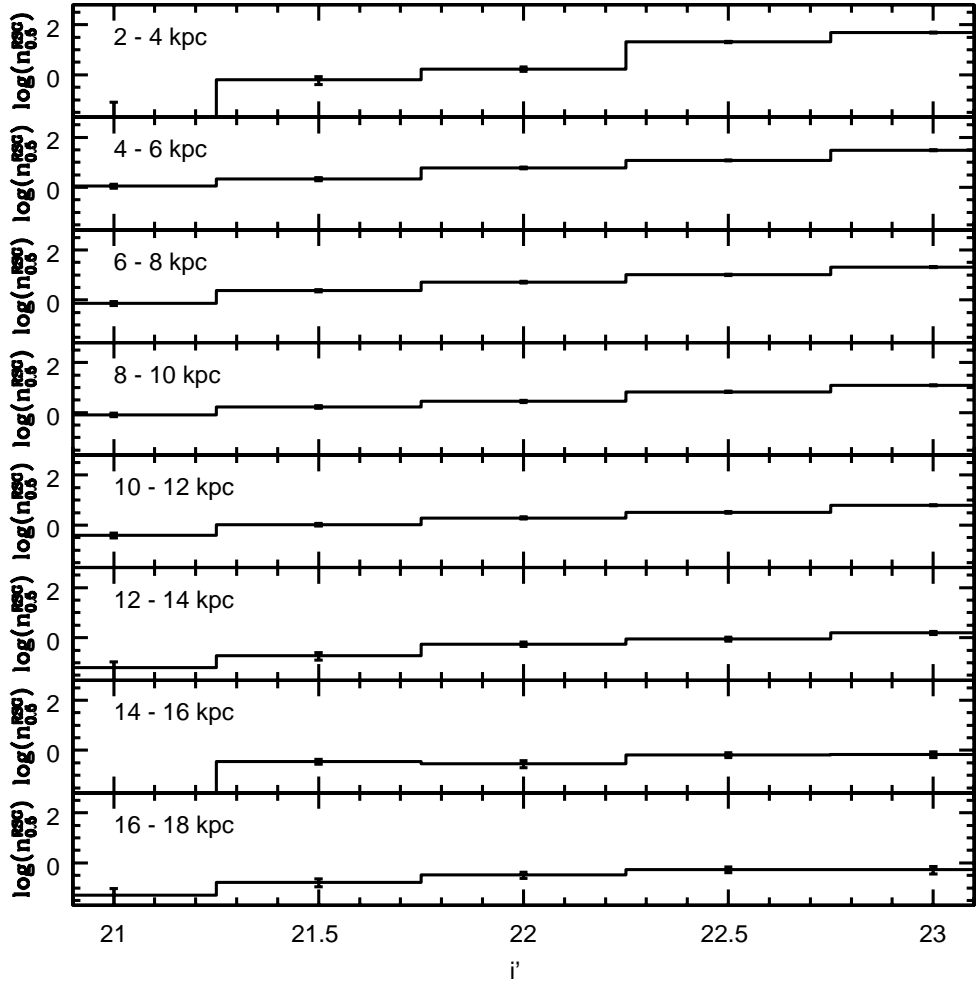}
\caption
{The same as Figure 10, but showing the LFs of stars that have $r'-i'$ between 
0.2 and 1.0, which are identified as RSGs.}
\end{figure}

\clearpage
\begin{figure}
\figurenum{12}
\epsscale{0.85}
\plotone{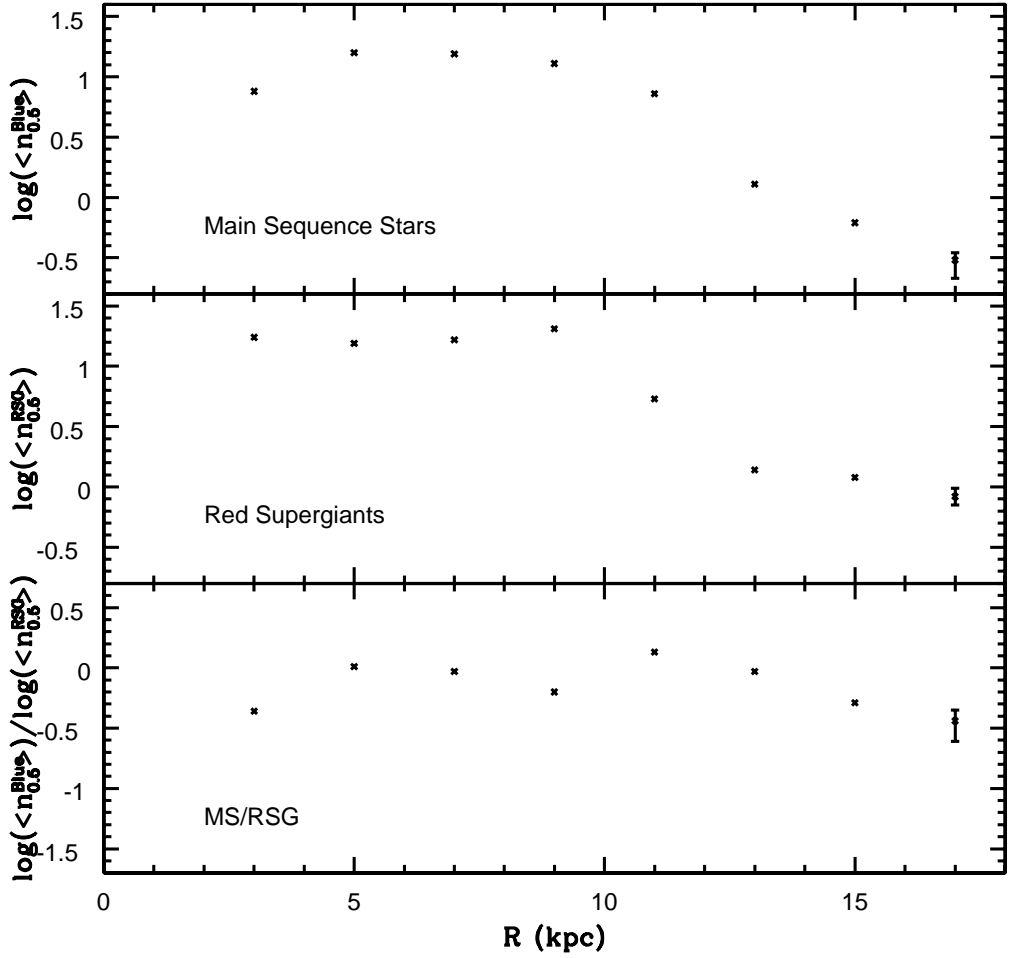}
\caption
{The radial behaviour of bright blue stars and RSGs, based 
on averages computed from the entries in Figures 10 and 11. 
$<n^{Blue}_{0.5}>$ is the mean number of objects per 0.5 magnitude interval per arcmin$^2$ 
in Figure 10 with $i'$ between 22.25 and 23.75, while $<n^{RSG}_{0.5}>$ is the mean 
number of objects per 0.5 magnitude interval per arcmin$^2$ in Figure 11 with $i'$ 
between 21.25 and 22.75. The ratio of the two means is shown in the bottom panel. 
The error bars for the 17 kpc data points reflect the random uncertainties in the number of 
sources in 16 -- 18 kpc annulus and in the control field. At smaller radii the error bars 
have a size that is comparable to, or smaller than, that of the plotted points, and 
hence are not shown.}
\end{figure}

\clearpage
\begin{figure}
\figurenum{13}
\epsscale{0.85}
\plotone{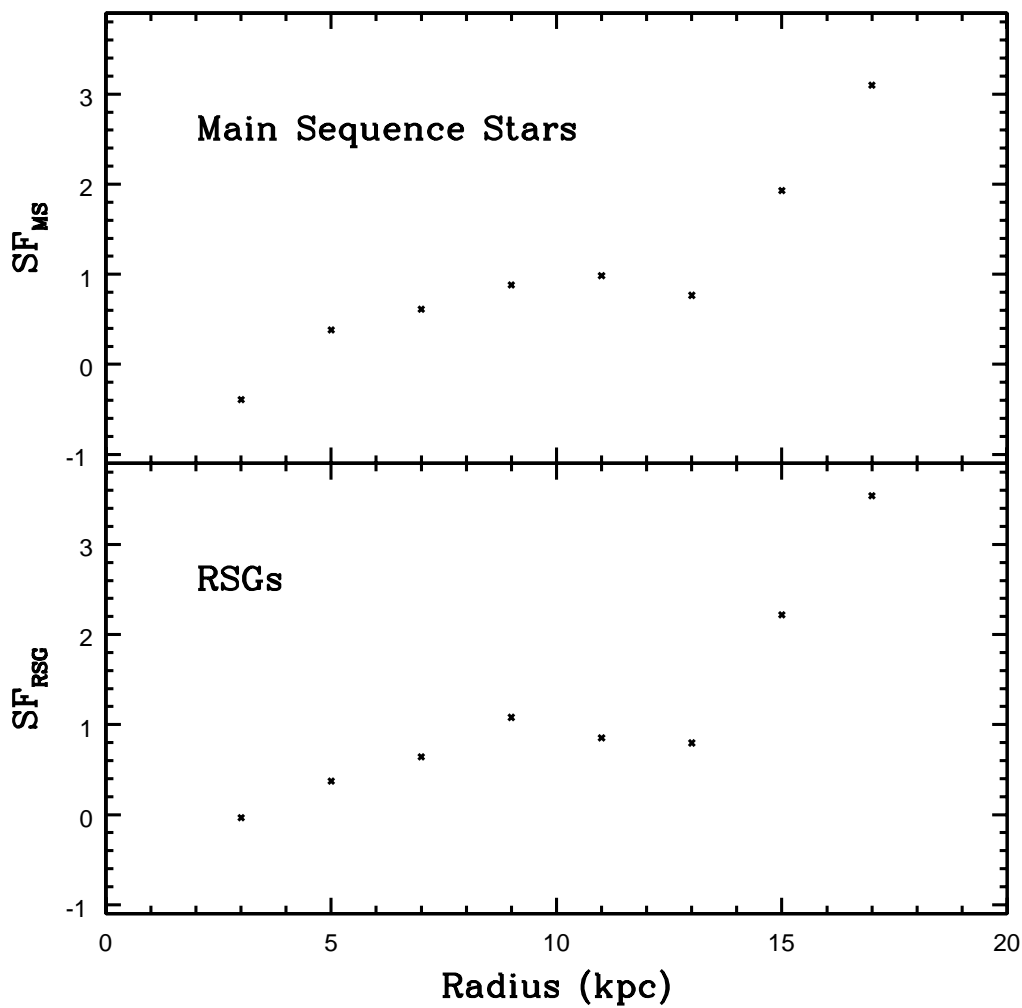}
\caption
{The specific frequencies (SFs) of bright blue stars and RSGs. The SF is defined as 
the number of stars in a system with an integrated brightness M$_K = -16$. 
Note that (1) there is a tendency for the SF of both bright blue stars and 
RSGs to increase with increasing R$_{GC}$, and (2) this trend steepens when 
R$_{GC} > 13$ kpc. The error bars due to random uncertainties in the numbers of sources in 
the various M81 annuli and the control fields are too small to be shown in this figure.
These data suggest that recently formed stars are not uniformly mixed 
throughout the M81 disk, but occur in progressively larger numbers as R$_{GC}$ increases.} 
\end{figure}

\clearpage
\begin{figure}
\figurenum{14}
\epsscale{0.85}
\plotone{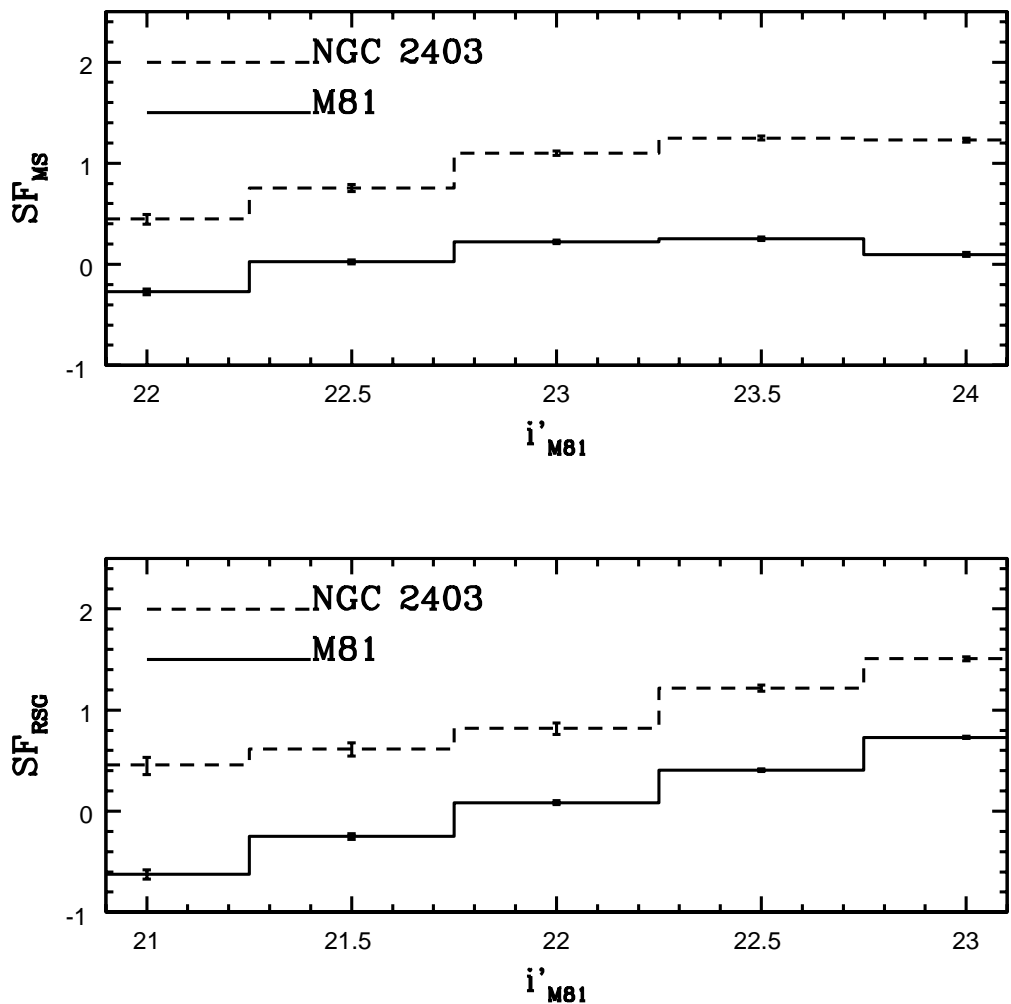}
\caption
{The SFs of bright blue stars and RSGs in M81 (solid lines) and NGC 2403 
(dashed lines). The M81 data is based on stars with R$_{GC}$ between 4 and 10 kpc, while the 
NGC 2403 data is for stars with R$_{GC}$ between 6 and 12 kpc. $i'_{M81}$ is the 
brightness that a star would have if viewed at the distance of M81. There is a 1.0 dex offset 
between the NGC 2403 and M81 SF curves, suggesting that the recent SSFRs 
in these galaxies differ by roughly a factor of ten.}
\end{figure}
 
\end{document}